\documentclass[12pt,preprint]{aastex}

\begin{document}
\title{The VLBA Imaging and Polarimetry Survey at 5 GHz}
\author{J. F. Helmboldt\altaffilmark{1}, G. B. Taylor\altaffilmark{1}, S. Tremblay\altaffilmark{1}, C. D. Fassnacht\altaffilmark{2}, R. C. Walker\altaffilmark{3}, S. T. Myers\altaffilmark{3}, L. O. Sjouwerman\altaffilmark{3}, T. J. Pearson\altaffilmark{4}, A. C. S. Readhead\altaffilmark{4}, L. Weintraub\altaffilmark{4}, N. Gehrels\altaffilmark{5}, R. W. Romani\altaffilmark{6}, S. Healey\altaffilmark{6}, P. F. Michelson\altaffilmark{6}, R. D. Blandford\altaffilmark{7}, and G. Cotter\altaffilmark{8}}

\altaffiltext{1}{Department of Physics and Astronomy, University of New Mexico, 800 Yale Blvd NE, Albuquerque, NM 87131, USA}
\altaffiltext{2}{Department of Physics, University of California at Davis, 1 Shields Avenue, Davis, CA 95616}
\altaffiltext{3}{National Radio Astronomy Observatory, P.O. Box O, Socorro, NM 87801, U.S.A.}
\altaffiltext{4}{Astronomy Department, California Institute of Technology, Mail Code 105-24, 1200 East California Boulevard, Pasadena, CA 91125}
\altaffiltext{5}{NASA Goddard Space Flight Center, Greenbelt, MD 20771}
\altaffiltext{6}{Department of Physics, Stanford University, Stanford, CA 94305}
\altaffiltext{7}{KIPAC, Stanford University, PO Box 20450, MS 29, Stanford, CA 94309, USA}
\altaffiltext{8}{University of Oxford, Department of Astrophysics, Denys Wilkinson Building, Keble Road, Oxford OX1 3RH}
\received{?}
\accepted{?}

\begin{abstract}
We present the first results of the VLBA Imaging and Polarimetry Survey (VIPS), a 5 GHz VLBI survey of 1,127 sources with flat radio spectra.  Through automated data reduction and imaging routines, we have produced publicly available I, Q, and U images and have detected polarized flux density from 37\% of the sources.  We have also developed an algorithm to use each source's I image to automatically classify it as a point-like source, a core-jet, a compact symmetric object (CSO) candidate, or a complex source.  Using data from the Sloan Digital Sky Survey (SDSS), we have found no significant trend between optical flux and 5 GHz flux density for any of the source categories.  Using the velocity width of the H$\beta$ emission line and the monochromatic luminosity at 5100 $\mbox{\AA}$ to estimate the central black hole mass, $M_{BH}$, we have found a weak trend between $M_{BH}$ and 5 GHz luminosity density for objects with SDSS spectra.  Ongoing optical follow-up for all VIPS sources will allow for more detailed explorations of these issues.  The mean ratio of the polarized to total 5 GHz flux density for VIPS sources with detected polarized flux density ranges from 1\% to 20\% with a median value of about 5\%.  This ratio is a factor of $\sim$3 larger if only the jet components of core-jet systems are considered and is noticeably higher for relatively large core-jet systems than for other source types, regardless of which components (i.e., core, jet, or both) are considered.  We have also found significant evidence that the directions of the jets in core-jet systems tend to be perpendicular to the electric vector position angles (EVPAs).  The data is consistent with a scenario in which $\sim$24\% of the polarized core-jets have EVPAs that are anti-aligned with the directions of their jet components and which have a substantial amount of Faraday rotation.  Follow-up observations at multiple frequencies will address this issue in more detail.  In addition to these initial results, plans for future follow-up observations are discussed.

\end{abstract}


\section{Introduction}
Very Long Baseline Interferometry (VLBI) can be a powerful tool for the detailed study of the nature of the centers of active galaxies.  Because VLBI can provide parsec-scale images at large distances and because active galactic nuclei (AGN) are among the few objects that have brightness temperatures that are high enough to be detected with VLBI instruments, VLBI and AGN are the perfect match of science and instrumentation.  Consequently, several VLBI surveys of AGN have been conducted \citep[e.g.,][]{pea88,tay96,kel98,bea02,lis05}.  These surveys have pioneered the effort to more fully understand the nature of AGN on parsec scales, but still do not provide both imaging and polarization data of samples large enough to answer many key questions such as:
\begin{itemize}
\item How do the nature and properties of jets evolve as they propagate from their black hole sources through different size scales out to the spatial extent of radio lobes (i.e., from $\sim$10 AU up to $\sim$10 kpc)?  Observations of AGN in the X-ray regime with the Chandra satellite will help explore this issue on relatively large ($\sim$1 kpc) scales.  Observations of emission within the gamma-ray regime conducted with the High Energy Stereoscopic System (HESS) have provided insights into the nature of the gamma-ray emitting regions ($\sim$0.01-1 pc) of AGN, but mostly for BL Lac objects \citep{aha05} .  A VLBI imaging survey combined with monitoring by the upcoming (launch in 2007) Gamma-Ray Large Area Space Telescope (GLAST) mission \citep{geh99}, and follow-up observations across the electromagnetic spectrum will be able to address this issue on relatively small ($\sim$10 pc) scales for a variety of AGN.
\item How are synchrotron radiation-emitting particles accelerated along jets and are these jets confined by toroidal magnetic fields or gas pressure?  A combination of radio spectral studies, multi-wavelength observations, and high spatial resolution polarimetric imaging would help answer these questions.  A VLBI survey with imaging and polarimetry for a large sample of AGN is the first key step in this process.
\item Is there a statistically significant trend between the direction of core magnetic fields and the direction of jets among different classes of AGN?  While such trends have been observed for quasars \citep{pol03} and BL Lacs \citep[e.g.,][]{gab00}, the lack of good polarization information for both classes of sources weakens the significance of these observed trends.  A relatively large sample of AGN with such polarization data would help this effort immensely.
\item How do radio sources associated with the central black holes of galaxies evolve and affect galaxy evolution?  A key to answering this question may be the subclass of objects known as Compact Symmetric Objects (CSOs) which may evolve into sources that resemble more well known classes of radio galaxies.  However, the relatively small number of known CSOs \citep{pec00} precludes any definite conclusion regarding their evolution.  A large sample of candidate CSOs imaged using VLBI at multiple epochs would help answer this question.
\end{itemize}
In order to address these key questions, we have compiled an imaging and polarimetry survey of 1,127 AGN with the NRAO Very Long Baseline Array (VLBA).  The VLBA Imaging and Polarimetry Survey (VIPS) consists of images and polarization data at 5 GHz with follow-up observations planned at different epochs and frequencies for selected sources.  Several VIPS sources will also be found to flare by GLAST, and will subsequently be the target of further VLBI follow-up campaigns to connect the presumed jet ejection with the gamma-ray flare.  This paper describes the sample selection and VLBA observations (\S 2.1), the data reduction and automated imaging (\S 2.2), and the classification of sources (\S 2.3).  A discussion of first results regarding the fraction of polarized flux density among the sources and a summary of future follow-up plans are contained in \S 3.\par

\section{Sample Definition and Data Processing}
\subsection{Sample Selection and Observations}
\begin{figure}
\plotone{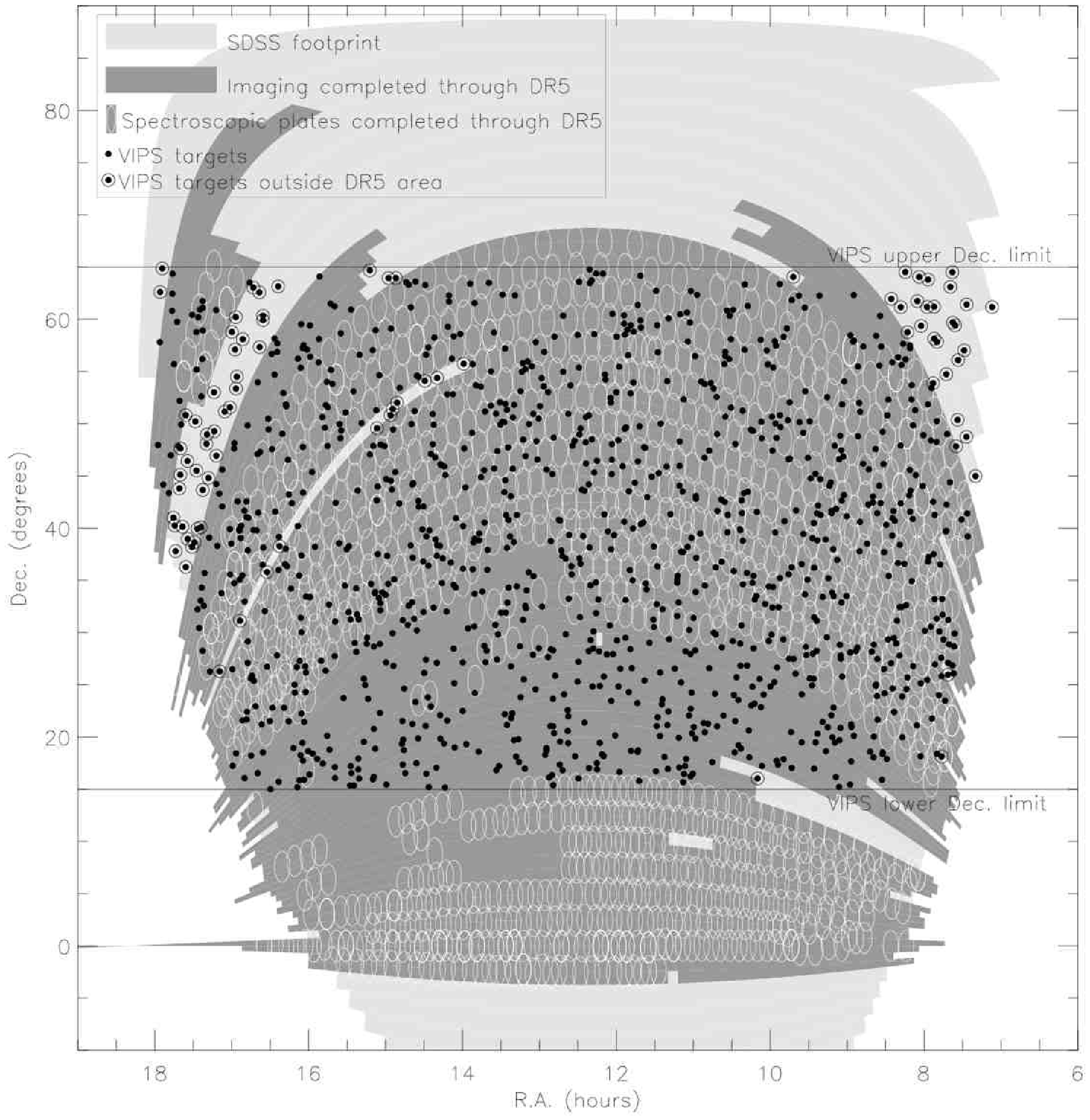}
\caption{The sky coverage of the VIPS source catalog superimposed on the sky coverage of the original SDSS footprint.  The light gray shaded area represents the original SDSS footprint; the dark gray shaded are represents the area with SDSS images through the fifth data release (DR5); the white open circles represent the spectroscopic plates observed through DR5.  The solid black points represent VIPS sources; the open black points represent those VIPS sources that are outside the SDSS DR5 imaging area.}
\label{sky}
\end{figure}

To meet the primary goals of this project, a relatively large sample of likely AGN, preferably with data from other wavelength regimes, is required.  To this end, we have have chosen the Cosmic Lens All-Sky Survey \citep[CLASS; ][]{mey03} as our parent sample.  CLASS is a VLA survey of $\sim$12,100 flat-spectrum objects ($\alpha > -0.5$ between 4.85 GHz and a lower frequency), making it an ideal source of likely AGN targets to be followed up with the VLBA.  We have also restricted our sample to lie on the survey area, or "footprint" of the Sloan Digital Sky Survey \citep[SDSS; ][]{yor00}.  Through the fifth data release of the SDSS \citep[DR5; ][]{ade06}, the imaging covers 8,000 square degrees and includes $\sim 2 \times 10^{8}$ objects.  Spectroscopy was obtained as part of the SDSS for $\sim 10^{6}$ of these objects, about $10^{5}$ of which are quasars.  We have chosen our source catalog so that all sources lie on the original SDSS footprint with an upper declination limit of 65$^{\circ}$ imposed to avoid the regions not imaged through DR5 (see Fig.\ \ref{sky}).  We also excluded sources below a declination of 15$^{\circ}$ because it is difficult to obtain good ($u,v$) coverage with the VLBA for these objects.  To keep the sample size large but manageable and to obtain a high detection rate without phase referencing, we selected all CLASS sources within this area on the sky with flux densities at 8.5 GHz greater than 85 mJy, yielding a sample of 1,127 sources.  Among these sources, 1,043 (93\%) have SDSS images and 356 (32\%) have SDSS spectra through DR5 with a median redshift of 1.2.\par
Among our target list, 141 have already been observed at 5 GHz with the VLBA as part of the Caltech-Jordell Bank Flat spectrum survey \citep[CJF; ][]{tay96,bri03,pol03}, 8 have been or will be observed at 15 GHz as part of the Monitoring of Jets in AGN with VLBA Experiments project \citep[MOJAVE; ][]{lis05}, and 20 were observed for the VIPS pilot program at 5 and 15 GHz \citep{tay05}.  Each of these surveys contain VLBA observations at 5 or 15 GHz that have adequate sensitivity for our purposes.  Both the MOJAVE and the VIPS pilot surveys were observed in full polarization, and \citet{pol03} obtained full polarization data for over half of the sources from the CJF survey that are also within our sample.  Because of this, we have elected to not re-observe sources from our sample contained within the CJF, MOJAVE, and VIPS pilot surveys.  The remaining 958 sources were observed with the VLBA within 18 separate observing runs of approximately 11 hours each from January to August 2006.  The targets for the runs consisted of groups of 52-54 VIPS catalog sources with four separate calibration sources, 3C279, J1310+3220, and some combination of DA193, OQ208, 3C273, J0854+2006, and J1159+2914.  Each VIPS target was observed for approximately 500 seconds divided into 10 separate scans.  All observations were conducted with four 8 MHz wide, full polarization IFs centered at frequencies of 4609, 4679, 4994, and 5095 MHz.  For these observations, an aggregate bit rate of 256 Mbps was used, yielding increased baseline sensitivity relative to the pilot program observations which used at a bandwidth of 128 Mbps.\par
All VLBA observations were scheduled using version 6.05 of the VLBA SCHED program.  Using built-in data regarding the locations and operation of the VLBA stations, a new mode in the updated version of SCHED will automatically produce a schedule for a list of targets with scan durations, a starting LST, and total experiment duration that is optimized both for ($u,v$) coverage and efficiency.  For each observing run, the starting LST and scan time per source was varied to produce a schedule that most efficiently used the entire duration of 11 hours while obtaining the vast majority (if not all) of the required scans for all targets.  Care was also taken to select the correct polarized calibration source(s) for each run so that it/they would be observed over a wide range of parallactic angle values while not significantly reducing the efficiency of the schedule for that run.

\subsection{Calibration and Source Mapping}
The bulk of the data calibration and imaging was performed in an automated way using both AIPS and DIFMAP scripts, similar to the process used in the VIPS pilot program.  The initial calibration and flagging of bad data was done using the VLBA data calibration pipeline \citep{sjo05} in AIPS.  All of the Stokes I, Q, and U images were generated using DIFMAP scripts which are described in detail in \citet{tay05}.  Polarized intensity and polarization angle images were made in AIPS.  For sources with data from only the CJF survey, the data was obtained from the NRAO archive and images were produced using the same procedure.  The basic procedure used is as follows:
\begin{enumerate}
\item The initial calibration is done with the VLBA pipeline.
\item Using this calibration, maps of the four calibration sources are made with the DIFMAP scripts.
\item In AIPS, the calibration is refined using the maps produced in step 2 as models for the calibration sources for self-calibration of the phases, and then again for self-calibration of both the phases and the amplitudes of the visibilities.  Polarization corrections are also determined using either DA193 or OQ208 and are applied to the data.
\item Using the new calibration, maps are made of all the sources with the DIFMAP scripts.  These maps are then used to identify the 16 brightest sources (excluding the calibration sources), which typically included all sources with peak flux densities $\sim$100 mJy or higher.  We chose to use the same number of calibration sources for each observing run rather than use all sources above a particular flux density limit so that the quality of the calibration would remain roughly constant among the runs.  Following the addition of new calibration sources, step 3 is then rerun using the maps of these sources for self-calibration (effectively yielding 20 calibration sources) to further refine the calibration solution.  This typically eliminates phase errors due to differences in right and left polarization and improves the amplitude calibration.
\item The position angle of 3C279 C4 is used to determine the corrections needed to align the observations at the four different frequencies.  After applying these corrections, maps of the calibration sources are remade with the DIFMAP scripts and the integrated polarization angle is measured for each of them using the Q and U maps for each of the four frequencies.  Observations of these sources from the 5 GHz VLA/VLBA polarization calibration database (see the acknowledgments) are then used to determine the mean phase correction needed to align the observed polarization angles with the true electric vector position angle (EVPA).
\item After applying the EVPA correction, final images are made of all sources with the DIFMAP scripts with a fixed restoring beam for all images.  For these maps, the minimum dynamic range required by the scripts for peaks to be identified within each image is lowered from 6 in previous applications to 5.5.  Q and U maps are made in all four frequencies as well as in pairs of frequencies, namely 4609 plus 4679 MHz and 4994 plus 5095 MHz.  This was done to improve the signal-to-noise ratio of the polarized flux density images (see the following step) by combining observations that are relatively close (within 100 MHz) of each other.  While not used here for polarization related measurements, the Q and U images made using each of the four different frequencies have been made publicly available (see below) for future applications.  It should be noted that the frequencies used do not provide enough separation in $\lambda^{2}$ for us to be able to obtain precise rotation measures (RMs) with which corrections for Faraday rotation may be applied.
\item The final I, Q, and U maps are used within AIPS to make images of both polarized flux density and polarization angle for the two pairs of frequencies using the COMP task including error biasing.  For any object that has polarized flux density that is significantly ($>5\sigma$) larger than the noise in its polarized flux density map, a contour map of the fraction of polarized flux density and polarization angle is made using the 4994 plus 5095 MHz pair.  It should be noted that no Stokes V images were created for our sources since most quasars do not have detectable amounts of circular polarization, and in instances where it is detected, the level of circular polarization is typically much lower than the level of linear polarization \citep[e.g.,][]{hom98}.
\end{enumerate}
\par
The typical noise measured from the 5 GHz I images, rms$_{\mbox{\scriptsize image}}$, is about 0.2 mJy beam$^{-1}$.  We have also computed the theoretical noise, rms$_{\mbox{\scriptsize theory}}$, according to
\begin{equation}
\mbox{rms}_{\mbox{\scriptsize theory}} = \frac{SEFD}{\eta_{s} \sqrt{N_{vis}\mbox{DR} \tau_{a}}}
\end{equation}
This equation holds for the simplified case of Stokes I images generated with $N_{vis}$ visibilities measured with identical antennas with natural weighting and no tapering, where $SEFD$ is the system temperature in units of Jy, $\eta_{s}$ is the system efficiency, DR is the data rate in bits s$^{-1}$, and $\tau_{a}$ is the time interval in seconds over which the visibilities were average to produce the $N_{vis}$ visibilities that were used to generate the image \citep{wro99}.  For the VLBA, the average value of $SEFD$ for the 10 antennas is about 300 Jy \citep{tay94}.  As mentioned earlier, for the newly observed VIPS sources, the data rate is 256 Mbps as compared to the data rate of 128 Mbps used for the pilot survey.  For those sources with data only from the CJF, the data rate ranges from 16 to 64 Mbps and was computed for each individual source assuming 1-bit sampling and using the total bandwidth and the number of independent polarizations used.  For the CJF data, we assumed a system efficiency of $1/\eta_{s}=1.8$ for the case of 1-bit sampling.   For the newly observed VIPS sources and those sources from the pilot survey, we assumed $1/\eta_{s}=1.84$ for the case of 2-bit sampling \citep{wro99}.\par
\begin{figure}
\plotone{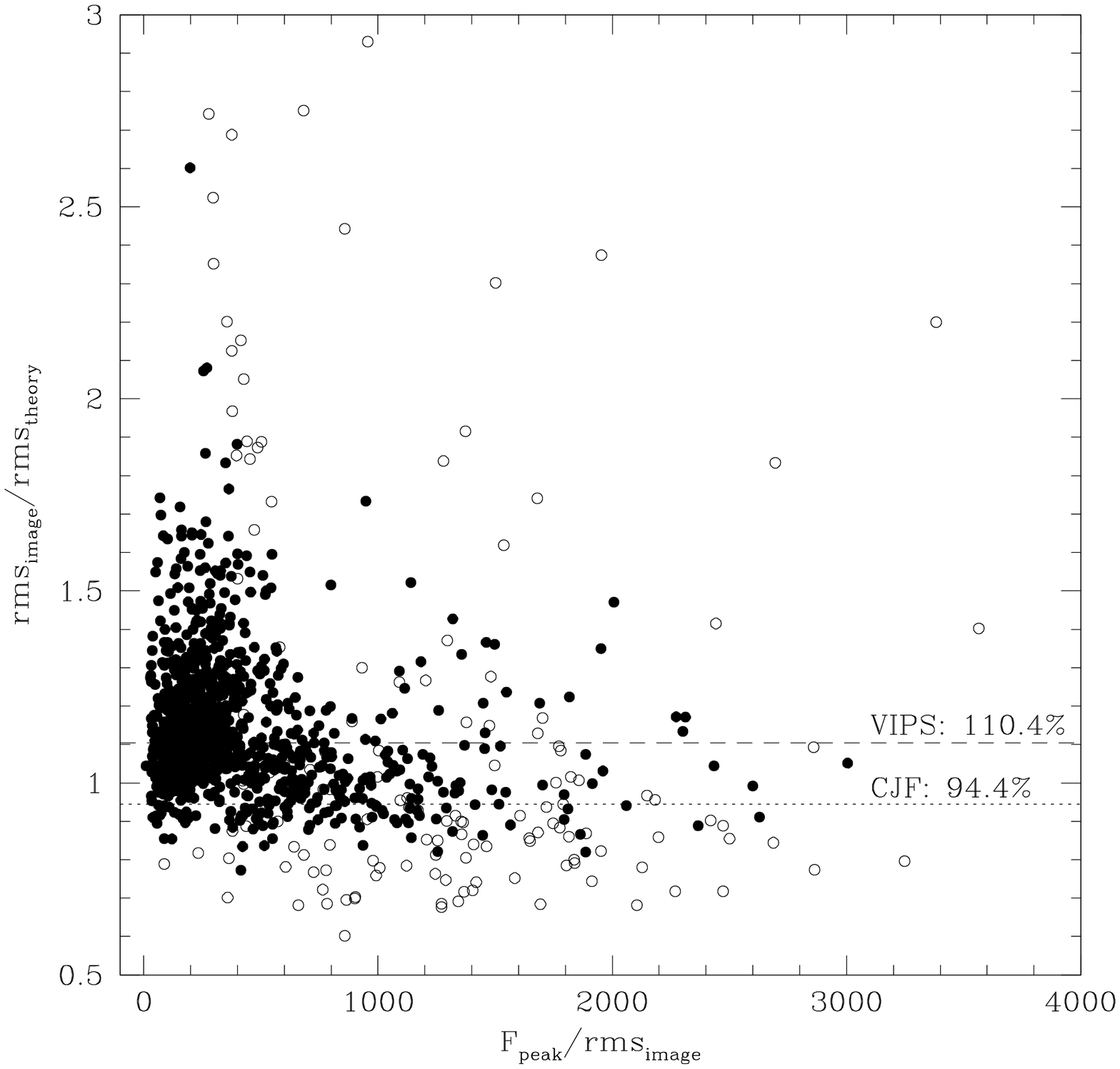}
\caption{The ratio of the rms noise measured from the 5 GHz image, rms$_{\mbox{\scriptsize image}}$, to the theoretical noise computed according to equation (1), rms$_{\mbox{\scriptsize theory}}$, for VIPS sources observed in 2006 (solid points) and VIPS sources imaged using data from the CJF survey \citep[open points, ][]{bri03} versus the peak signal-to-noise ratio from the 5 GHz image.  The median ratios of rms$_{\mbox{\scriptsize image}}$/rms$_{\mbox{\scriptsize theory}}$ are represented by a dashed line for the VIPS sources observed in 2006 and by a dotted line for CJF/VIPS sources.}
\label{noise}
\end{figure}

In Fig.\ \ref{noise}, we have plotted the ratio of rms$_{\mbox{\scriptsize image}}$ to rms$_{\mbox{\scriptsize theory}}$ versus the peak signal-to-noise ratios for the images.  More than 80\% of the newly observed VIPS sources have rms$_{\mbox{\scriptsize image}}>$rms$_{\mbox{\scriptsize theory}}$ and nearly all of these sources have rms$_{\mbox{\scriptsize image}}<1.7\;$rms$_{\mbox{\scriptsize theory}}$.  This implies that in general, the newly acquired data is not dynamic range limited and that our automated use of the CLEAN algorithm has not artificially reduced the noise level of the images.  For the data taken from the CJF survey, however, more than half of the sources have rms$_{\mbox{\scriptsize image}}<$rms$_{\mbox{\scriptsize theory}}$.  While our simplified computation of rms$_{\mbox{\scriptsize theory}}$ has not taken into account factors such as weather and elevation, a proper treatment of such effects would most likely increase the values of rms$_{\mbox{\scriptsize theory}}$ and worsen the discrepancy between rms$_{\mbox{\scriptsize image}}$ and rms$_{\mbox{\scriptsize theory}}$.  We believe that the quality of the CJF data has primarily contributed to the relatively low values of rms$_{\mbox{\scriptsize image}}$ as the images produced from these data can have considerable peaks within the noise which may lead to the images being "over-cleaned" by the automatic imaging process, thus artificially reducing the rms noise of the resulting images.  We therefore caution that the reader that the noise measured from the images for sources with data only from CJF, 141 sources in all, may underestimate the true noise.  One may identify these sources in Table \ref{props} by their observing dates (i.e., all sources observed before 2004).\par
In a few ($<$1\%) cases, the final 5 GHz images were of relatively poor quality and were remade within DIFMAP "by hand".  The poor quality of the model fits for some of these cases was found to be largely due to significant flux density that was outside of the original image in which case larger maps were made.  Because the self-calibration process used to make the images can produce false point-like sources from peaks in the noise, especially for faint objects that are essentially not detected, the maps and visibility data for all sources with peak flux densities $<$20 mJy beam$^{-1}$ were inspected, 27 objects in all.  The reliability of the maps for these sources was evaluated on a case-by-case basis.  The images for a total of eleven sources were deemed unreliable and flagged as non-detections and make up about 1\% of the sample.  For all detected sources, the total and peak flux densities at 5 GHz are listed in Table \ref{props}, including those sources observed as part of the CJF and VIPS pilot surveys.  The 8 sources from the full catalog of 1,127 sources that have or will have data only from the MOJAVE survey are not included.  The full version of Table \ref{props} is available in electronic form only.  The version displayed here merely illustrates the general format of the table.  All Stokes I, Q, and U maps are publicly available in FITS format via the VIPS website, \mbox{http://www.phys.unm.edu/$\sim$gbtaylor/VIPS}, as well as the VLBA ($u,v$) data (also in FITS format) and PDF and GIF versions of contour plots of the I maps, polarization contours, and visibility plots.  The full VIPS source catalog (i.e., including the 8 MOJAVE sources excluded from Table \ref{props}) is also available via the VIPS website.

\subsection{Source Classification}
\begin{figure}
\plotone{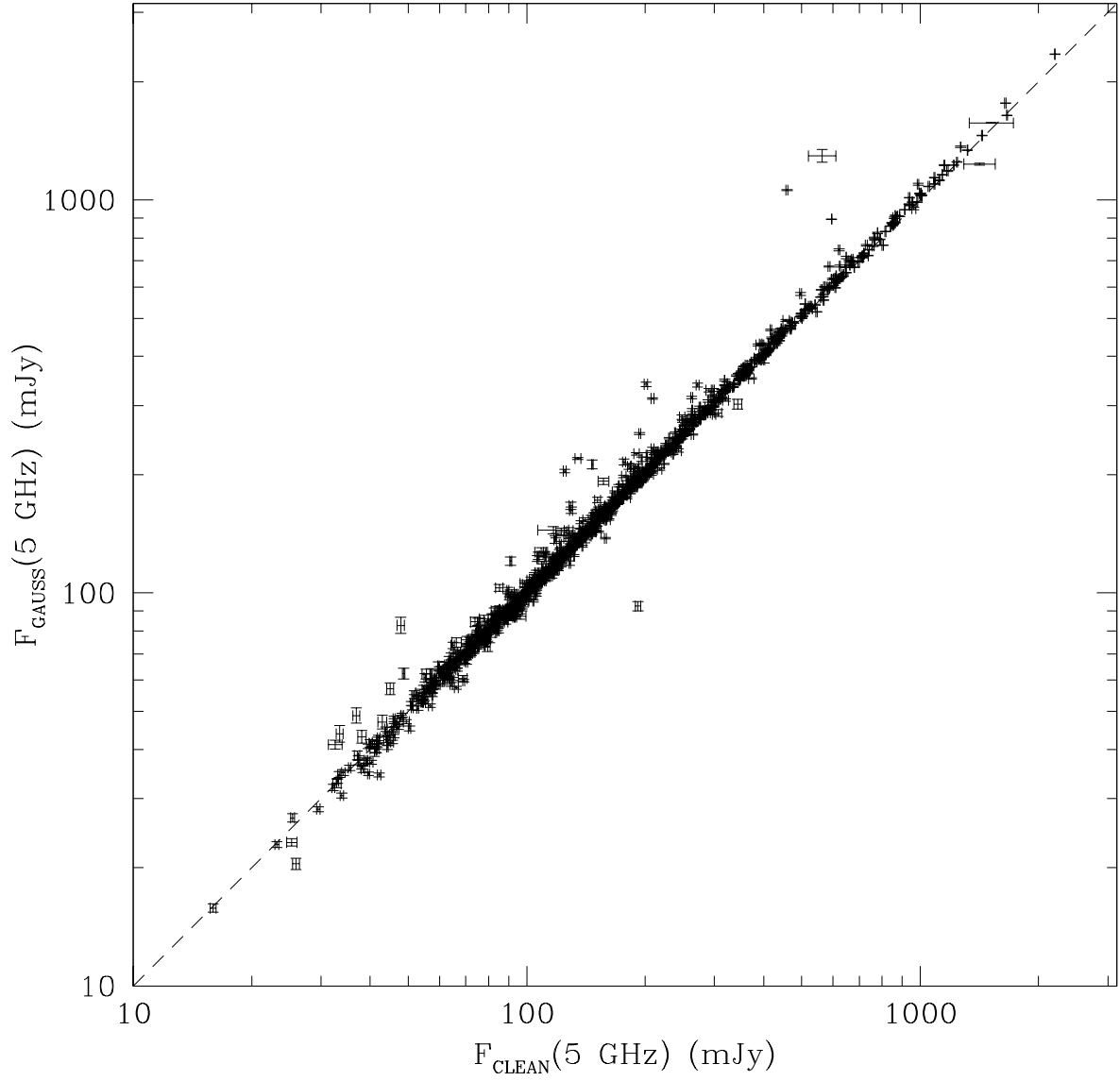}
\caption{The total flux density at 5 GHz contained within the components from the Gaussian fits (see \S 2.2) versus the total cleaned flux density.  The dashed line is not a fit, but simply the line expected for $F_{GAUSS}$=$F_{CLEAN}$ plotted for reference.}
\label{fcfg}
\end{figure}

The relatively large sample size implies that the VIPS catalog will contain a variety of source types.  In order to provide some level of information regarding the morphology of the VIPS sources, we have divided the sources into four categories (1) point-like sources (PS), (2) core-jets, which are subdivided into short jets (SJET) and long jets (LJET), (3) compact symmetric object candidates (CSO), and (4) complex sources (CPLX).  To deal with the relatively large number of sources and to make the classifications as objective as possible, we have developed an automated classification procedure based on Gaussian component fitting performed within the image plane.  The Gaussian fitting was performed within AIPS with the SAD task which fits elliptical Gaussians to all sources within an image down to a given limiting flux density.  Through trial and error, it was found that the best results were obtained if the Gaussian fits were performed for sources at ten different flux density limits starting with the peak flux density for the image and proceeding down to the 6$\sigma$ level in steps evenly spaced in logarithmic flux density.  Examples of the Gaussian fits displayed in Fig.\ \ref{exgauss} illustrate that the fitting reliably breaks up the objects into distinct components.  In Fig.\ \ref{fcfg}, we plot the total flux density contained within the Gaussian components versus the total cleaned flux density for all detected sources.  While the Gaussian fits clearly overestimate the flux density in some instances, in general, the flux density of the Gaussian components matches the cleaned flux density relatively well.\par
To classify the VIPS sources as objectively as possible, we have developed the following algorithm which utilizes the Gaussian components:
\begin{enumerate}
\item If a source has one Gaussian component that contains 95\% or more of the total flux density of all its Gaussian components, it is flagged as a single component object.  Single component objects that are more elongated than the restoring beam used (i.e., an axis ratio of $b/a<0.6$), are classified as core-jets.  Those sources not classified as core-jets are classified as point-like (PS).
\item Sources not flagged as single component sources are flagged as double sources if their two brightest Gaussian components contain 95\% or more of the total flux density.  If the flux densities and fitted sizes (assumed to be proportional to $ab$) of these components agree within a factor of 2.5, they are classified as compact symmetric object candidates (CSO).  If this is not the case, they are classified as core-jets.
\item Sources not flagged as single or double sources are flagged as multiple component sources.  The dominant components within each of these sources are identified as the brightest sources whose combined flux density is 95\% or more of the total flux density.  For these dominant sources, a line is fit to their relative declinations as a function of their relative right ascensions.  If the dispersion of the positions of the components relative to their center (taken to be the mean position of the components) in the direction of this fitted line is a factor of two greater than the dispersion perpendicular to the fitted line, the object is classified as a core-jet.  Otherwise, the object is classified as complex (CPLX).
\item For all sources classified as core-jets, those longer than 6 mas are classified as long jets (LJET).  Those shorter than this limit are classified as short jets (SJET) according to \citet{pol03} who found that there may be difference between the polarization properties of quasars with jets that are divided into two groups using this limit.
\end{enumerate}
Following this initial classification, an additional algorithm was run to perform a more detailed search for CSO candidates with morphologies that are more complex than symmetric double sources.  This algorithm first identifies sources with two Gaussian components whose combined flux density is greater than 80\% of the total flux density and whose flux densities and fitted sizes ($\propto ab$) agree within a factor of 2.5.  This was done to include double sources that may have some extended emission that could still be CSOs.  Any multiple component object classified as LJET with a total length greater than 12 mas whose brightest Gaussian component was within 3 mas of the mean position of all components was also reclassified as a CSO.  This was done to include any source that appears to have a core with significant and roughly symmetric diffuse emission on opposite sides of the core oriented along a single axis.\par
\begin{figure}
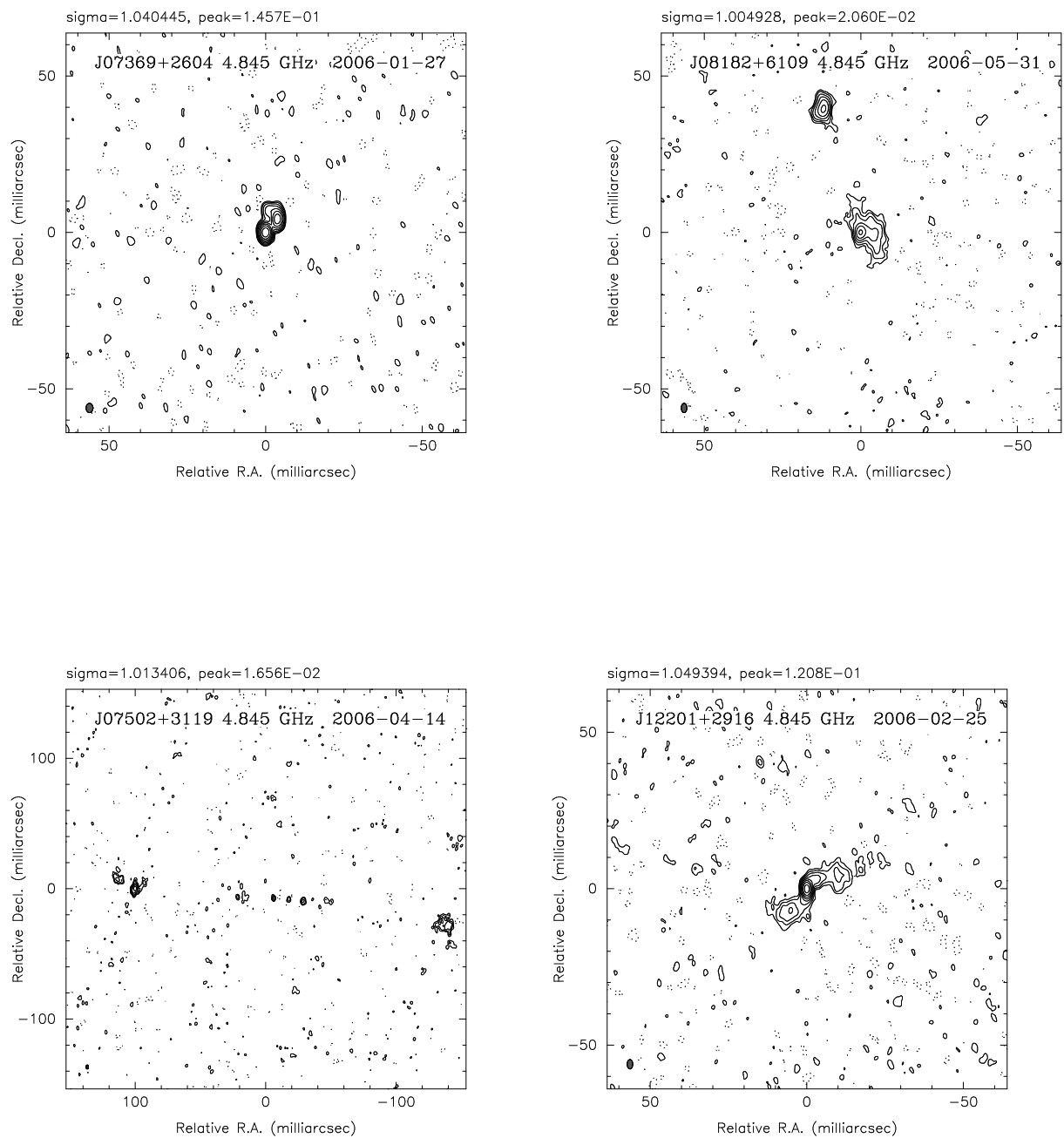

\plottwo{f7a.eps}{f7b.eps}
\plottwo{f7c.eps}{f7d.eps}
\caption{Examples of four sources reclassified as CSO candidates by the additional specialized CSO classification algorithm (see \S 2.3).    Here, "sigma" and "peak" refer to the reduced $\chi^{2}$ between the CLEAN components and the data and the peak cleaned flux density in Jy beam$^{-1}$ respectively.}
\label{csos}
\end{figure}

Following this, the algorithm identifies groups of Gaussian components that overlap on the image where in practice, the "groups" are allowed to have as little as one component.  For this purpose, the outer boundary of each component was defined to be an ellipse with the same position angle as the Gaussian component and major and minor axes equal to $3a/\sqrt{8\mbox{ln}2}$ and $3b/\sqrt{8\mbox{ln}2}$ where $a$ and $b$ and the full widths at half maximum of the Gaussian component along the major and minor axes respectively (i.e., effectively 3$\sigma$ from the center of the Gaussian component).  Components that had outer boundaries that were defined in this manner which intersected were considered to be overlapping components.  Using these groups, the algorithm identifies the following objects as CSO candidates:  (1) objects with two groups that contain 80\% of the total flux density and whose flux densities agree within a factor of 2.5, (2) objects that have at least two multiple component groups where the group closest to the center (equal to the mean component position) has only one component (i.e., it is likely the core at the center of the CSO), and (3) objects with more than two groups where the brightest component is closest to the center.  Images of examples of sources reclassified as CSO candidates by this algorithm are displayed in Fig.\ \ref{csos}.\par
\begin{figure}
\plotone{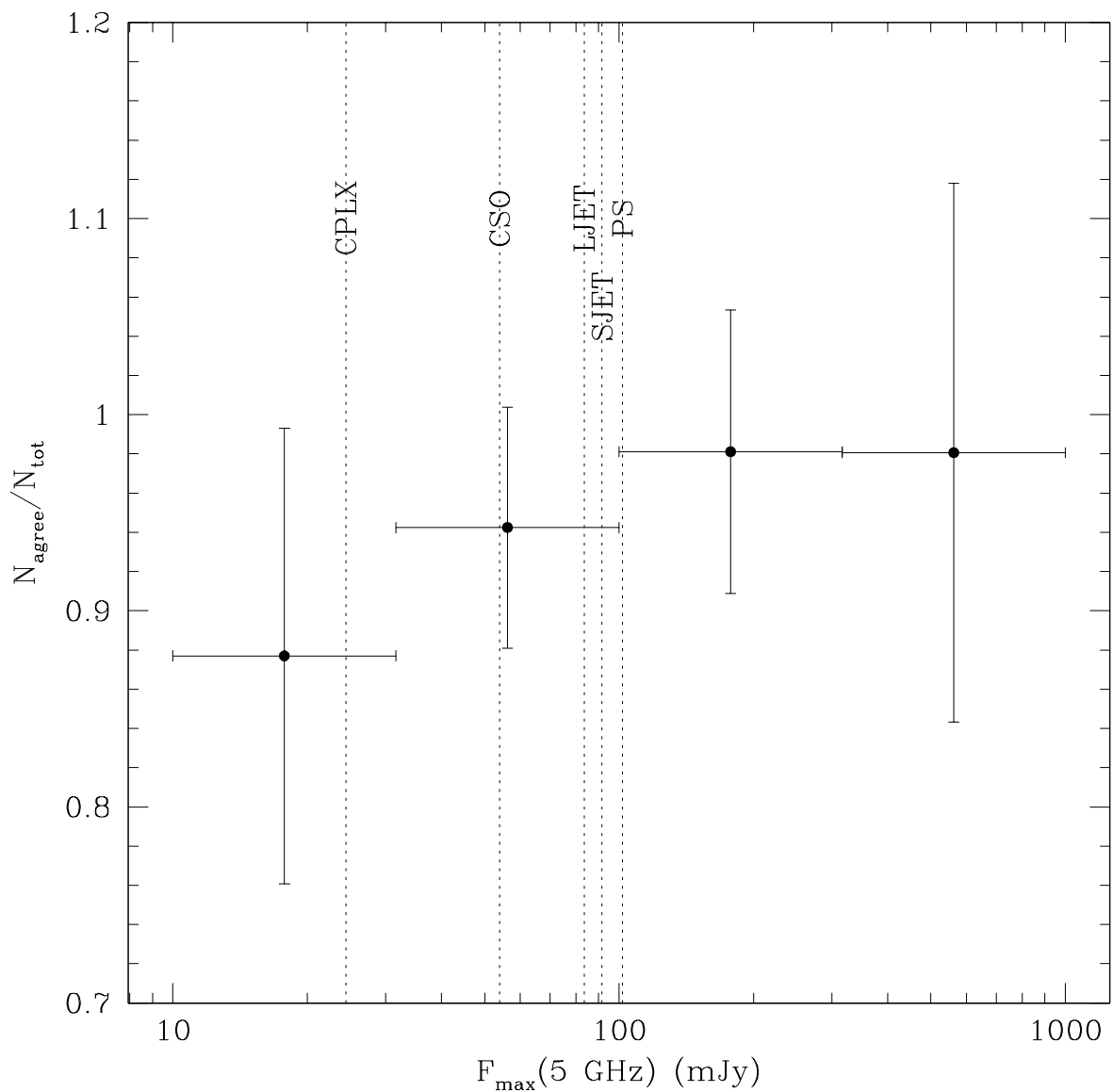}
\caption{Within bins of peak 5 GHz flux density, the fraction of sources for which the ``bye-eye'' classification agrees with the automatic classification algorithm (see \S 2.3).  The median peak 5 GHz flux densities for the five sources categories are displayed as vertical dotted lines.}
\label{2class}
\end{figure}

To test the quality of the automatic classifications, visual inspection of the I image for each source was performed.  The "by-eye" classifications derived from these visual inspections agreed with the automatic classifications in 99\% of the cases for both point-like objects and short jets, in 93\% of the cases for long jets, in 87\% of the cases for CSO candidates, and in 71\% of the cases for complex objects.  The larger discrepancy between the two classifications for CSO candidates and complex objects appears to have more to do with peak flux density than with morphology.  In general, the by-eye classifications agree with the automatic classifications more frequently for brighter objects.  This is illustrated in Fig.\ \ref{2class} where we have plotted the fraction of sources for which the two classifications agree within bins of peak 5 GHz flux density.  These results show that for sources with peak flux densities greater than $\sim$60 mJy, the two classifications agree in 95\% of the cases.  The median peak flux densities for the CPLX, CSO, LJET, SJET, and PS sources are 24, 54, 84, 92, and 102 mJy beam$^{-1}$ respectively, implying that higher fractions of CSO and CPLX sources are misclassified due to their relatively low flux densities and not their complex morphologies.\par
Based on the comparison with the by-eye classifications, the automatic classification algorithm appears to provide reliable and objective source types and may be used successfully with follow-up VLBI observations of the VIPS sample or with other VLBI imaging surveys.  However, since the performance of the algorithm is lower for the typically fainter CSO candidates and complex sources which are among the rarest and most interesting sources in the sample, we have elected to make our by-eye classifications available as well to facilitate follow-up observations of these types of sources.  To this end, we have listed the automatically determined source types in Table \ref{props} along with the by-eye classifications for those sources where the two classifications disagreed.  For objects flagged as non-detections, the type is listed as ND.  A summary of the number of sources in each category as well as the fraction of sources of each type with detected polarized flux density is contained in Table \ref{classes}.  For Table \ref{classes} and the remainder of the paper, the source types used include the by-eye re-classifications where applicable.\par
For each object, the number of dominant Gaussian components is listed in Table \ref{props}.  For each object, these dominant Gaussian components were also used to measure a radius equal to $\sqrt{ab}$ for single component sources and equal to the average distance of the components from their centroid for double and multiple component sources which is listed in Table \ref{props}.  For double and multiple sources, the maximum separation among the components of each object is also listed in Table \ref{props} along with a position angle for any objects classified as core-jets computed using a linear fit to the relative right ascensions and declinations of the dominant Gaussian components.

\subsection{Polarization Properties}
To exploit the relatively large number of source with full polarization data provided by VIPS, we have developed the following automated method for measuring the polarization properties of different components of each source.  First, for each source with detected ($>5\sigma$) polarized flux density, we constructed three image masks, one using all the Gaussian components fit to the I image (see \S 2.3), one using only the brightest Gaussian component, and one that is the difference between these two masks.  We then constructed another image mask using the polarized intensity and noise images produced by the AIPS task COMB using the Q and U images from the 4994 plus 5095 MHz pair (see \S 2.2) including error biasing.  This mask was made by setting pixels with signal-to-noise ratios $>$5 to unity and the remaining pixels to zero.  We then constructed three composite masks by multiplying the polarized intensity mask by the three masks made using the Gaussian components and used them with the polarized flux density and I images to measure the mean polarization fraction, $f_{pol}$, or the ratio of polarized to total intensity.  Using the three separate masks, we obtained measurements of  $f_{pol}$ for the entire object, the object's core, which we assumed to be represented by the brightest Gaussian component, and the regions outside the core.  In the majority ($\sim$55\%) of sources with more than one dominant Gaussian component, the polarized flux density is only found within the core.  About 35\% of these sources have detected polarized flux density both within and outside the core while about $\sim$10\% have detected polarized intensity only in the regions outside the core.\par
We also used the same three composite masks to measure a polarized intensity-weighted mean EVPA for the whole object, the core, and the regions outside the core using the polarized flux density image and the polarization angle image, also produced using the COMB procedure.  For sources with both positive and negative EVPA values in their polarization angle images, care was taken to ensure that the mean EVPA was computed properly to ensure that, for instance, for a source with EVPA values near both 90$^{\circ}$ and -90$^{\circ}$, the mean EVPA was near either 90$^{\circ}$ or -90$^{\circ}$ and not 0$^{\circ}$.  Specifically, in each of these instances, mean EVPA values were computed separately for the positive and negative pixels on the polarization angle image.  If the difference between these two mean values was less than 90$^{\circ}$, a polarized intensity-weighted EVPA was computed using all the pixels from the polarization angle image.  If the two mean values differed by more than 90$^{\circ}$, 180$^{\circ}$ was added to each negative pixel on the polarized image after which the polarized intensity-weight mean EVPA was computed.  If this mean EVPA was greater than 90$^{\circ}$, 180$^{\circ}$ was subtracted from it so that all of the polarized intensity-weighted EVPA values would be between $-$90$^{\circ}$ and 90$^{\circ}$.  All polarization related quantities are listed in Table \ref{polprops} for all objects with detected polarized flux density.  As with Table \ref{props}, we have included only the first twenty sources here to provide an example of the table format, and the entire table is available in electronic form only.  We have used the flux densities from the I, Q, U, and polarized intensity images along with the rms values measured from the I, Q, and U images to estimate the uncertainties in both $f_{pol}$ and $\chi$ and find that the typical errors in these quantities are 0.003 and 3$^{\circ}$ respectively.

\section{Results and Future Work}
The compilation of images and polarization data that make up VIPS constitute the largest such database of AGN to date.  A number of scientific endeavors to explore the nature of AGN are possible with these data.  Here, we will briefly explore the most basic properties of the sample and will leave more detailed analysis for subsequent papers.
\subsection{Comparison with Optical Data}
\begin{figure}
\plotone{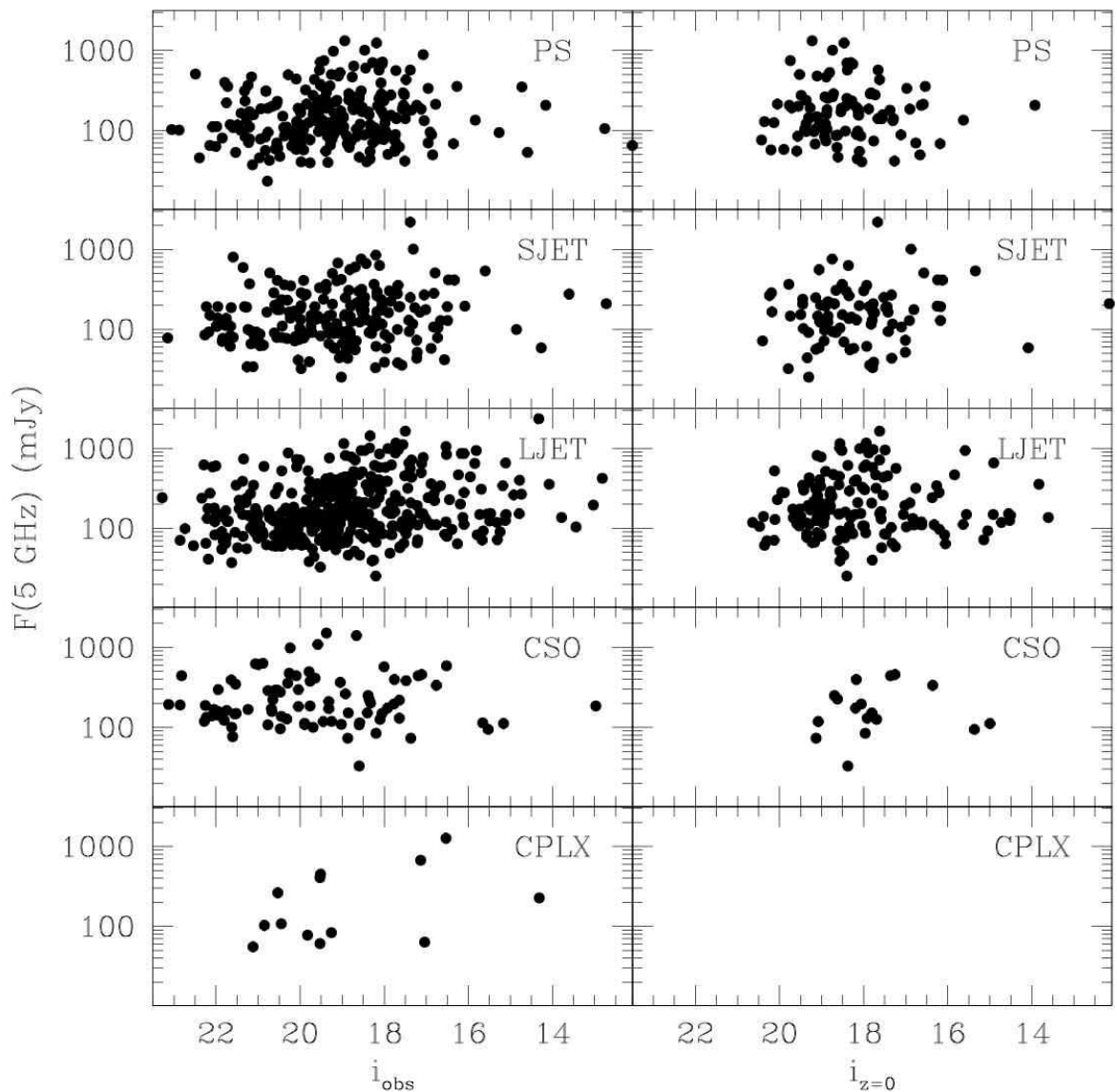}
\caption{The 5 GHz flux density measured from the VIPS images versus the SDSS $i$-band magnitude for each of the five source categories (see \S 2.3).  In the left panels, the observed $i$-band SDSS magnitudes are used; in the right panels, only the sources with SDSS spectra are included and their $i$-band magnitudes have been K-corrected to z=0 using the corrections of \citet{ric06}.}
\label{sdss}
\end{figure}

Since the VIPS sample was chosen to lie on the SDSS survey footprint, there are existing optical data for nearly all of our sources.  Through DR5, 997 (88\%) have optical magnitudes measured by the SDSS photometric pipeline \citep[see ][]{sto02} and 356 (32\%) have SDSS spectra.  In Fig.\ \ref{sdss}, we have plotted the 5 GHz flux densities measured from the VIPS images versus the SDSS $i$-band magnitudes separately for each of the five source categories.  For those VIPS sources with SDSS spectra, we have used the sources' redshifts and the K-corrections of \citet{ric06} to correct their $i$-band magnitudes to a redshift of z=0 and have plotted these sources separately in the right panels of Fig.\ \ref{sdss}.  For both the observed and K-corrected $i$-band magnitudes, there is no clear trend between the 5 GHz flux densities and optical magnitudes for any of the five source categories.  This illustrates the need for further optical follow-up observations to obtain redshifts for as many VIPS sources as possible so that consistent distance measurements may be obtained without any bias introduced by the selection function of the SDSS and any additional optical selection effects.  Optical spectra are currently being obtained for all VIPS sources not targeted for spectroscopy by the SDSS.  A detailed discussion of the results of this work will be presented in a subsequent paper.\par
One of the most useful properties of broad-line AGN that can be estimated using optical spectra is the virial mass of the central black hole, assumed to be equal to $G^{-1}R_{BLR}V^{2}$ where $R_{BLR}$ is the radius of the broad line region and $V$ is the velocity width at half maximum of the broad optical emission lines.  \citet{kas00} have demonstrated that $R_{BLR}$ is strongly correlated with the monochromatic continuum luminosity at 5100 $\mbox{\AA}$, $L_{5100}$.  Using their observed correlation, one can use the velocity width of the H$\beta$ emission line, $V(H\beta)$, with $L_{5100}$ to estimate the central black hole mass, $M_{BH}$.  With such a large sample of high resolution radio frequency images of sources that have or will have optical spectra, we are in a good position to explore any relation between $M_{BH}$ and the radio frequency luminosity of the cores of AGN or other properties.\par
\begin{figure}
\plotone{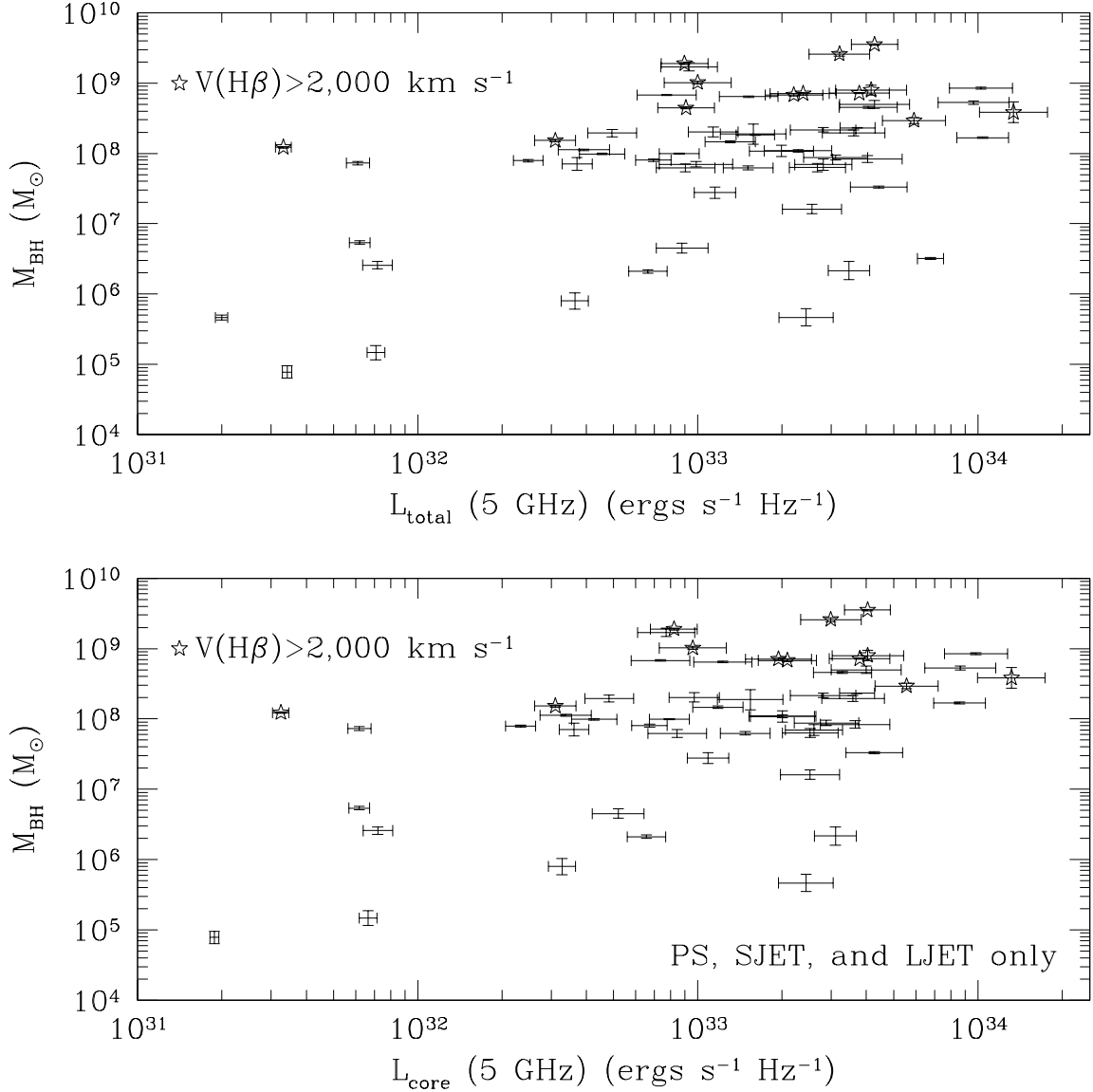}
\caption{For sources with SDSS spectra and $>3\sigma$ detections of the H$\beta$ emission line, the estimated viral mass of the central black hole (see \S 3) versus the total 5 GHz luminosity density (upper) and the luminosity density of the core component for point-like and core-jet sources (lower) taken to be the luminosity density of the brightest Gaussian component (see \S 2.3).  In both panels, the error in each luminosity density reflects the range in rest-frame 5 GHz luminosity densities expected for a power-law slope of $-0.5 < \alpha < 0.5$.  The Spearman rank-order correlation coefficient between $M_{BH}$ and $L(\mbox{5 GHz})$ is 0.4.  The probability of getting this result by chance is $\sim$50\%, implying that the observed trend is rather weak.  Similar results were obtained when only sources with H$\beta$ velocity widths, $V(H\beta)$, greater than 2,000 km s$^{-1}$ (represented as stars in both panels) were considered and when only $L_{core}(\mbox{5 GHz})$ was considered for point-like and core-jet sources.}
\label{wvel}
\end{figure}

Using the Gaussian fits to the emission lines performed by the SDSS spectroscopic pipeline \citep[see ][]{sto02}, we have computed $V(H\beta)$ for all sources that have $>3\sigma$ detections of the H$\beta$ emission line and which are at low enough redshifts that values for rest-frame $L_{5100}$ could be determined, 62 sources in all.  In Fig.\ \ref{wvel}, we have plotted the estimated values of $M_{BH}$ for these 62 sources versus their total 5 GHz luminosity densities and versus the luminosity density of the core component of each point-like and core-jet source (57 sources in all), which we take to be the luminosity density of the brightest Gaussian component.  In both cases, we have included errors in the luminosity densities that reflect the range in rest frame 5 GHz luminosity density expected for power-law spectra with slopes ranging from $-$0.5 to 0.5.  We find a slight correlation between black hole mass and 5 GHz luminosity density with a Spearman rank-order correlation coefficient of 0.4.  The probability of getting this result by chance, however, is about 50\%, indicating that the trend is weak at best.  The results are nearly the same if we only consider sources with $V(H\beta)>$2,000 km s$^{-1}$, or if we only consider the core luminosity for point-like and core-jet sources.  However, with the completion of the optical follow-up, we will be able to explore this issue much more thoroughly using velocity widths and radio luminosities for $\sim$3 times as many sources.
\subsection{Polarization Results}
\subsubsection{Fractional Polarization}
\begin{figure}
\plotone{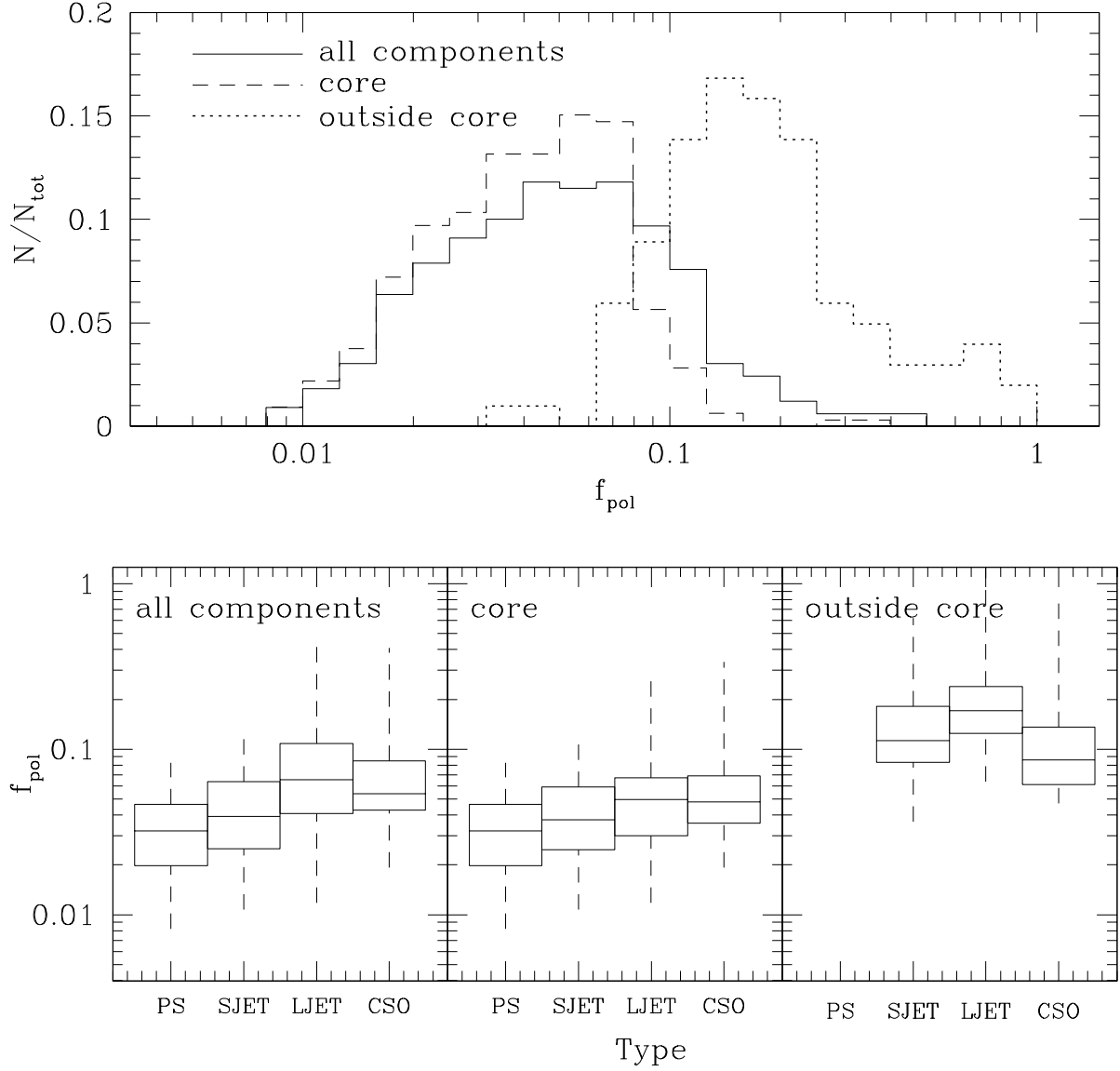}
\caption{Upper:  The distribution of the mean 5 GHz polarization fraction, $f_{pol}$, measured for the entire object (solid line), the core (i.e., the brightest Gaussian component; dashed line), and the regions outside the core (dotted line; see \S 2.4).  Lower:  a so called "box-and-whisker" representation of the three $f_{pol}$ distributions for each source category (see \S 2.3) excluding complex sources for which there were only two sources with detected polarized intensity.  The lower and upper boundaries of each box represent the 25$^{\mbox{\scriptsize th}}$ and 75$^{\mbox{\scriptsize th}}$ percentiles respectively.  The horizontal line within each box represents the median, and the dashed lines extend to the extrema of each distribution.}
\label{pfrac}
\end{figure}

Perhaps the most distinguishing aspect of VIPS is the number of sources with detected polarized flux density, 393 sources in all and about 37\% of all newly observed sources (i.e., excluding sources with data only from the CJF or VIPS pilot surveys).  This puts us in a position to be able to accurately measure, among a relatively large sample of AGN, the distribution of fractional polarization, $f_{pol}$, for different source components (see \S 2.4) and different source types.  In Fig.\ \ref{pfrac}, we have plotted the $f_{pol}$ distributions for all newly observed sources for the entire objects, the objects' cores, and the regions outside the cores.  Using all regions of polarized intensity from each object, the median value of $f_{pol}$ is about 5\% and ranges from 1-20\%.  The results are similar if only the cores are used.  However, $f_{pol}$ is significantly larger on average for regions outside the cores with a median value of $\sim$17\% and a range of about 7-100\%.  These regions are generally the jets of core-jet systems since 90\% of systems with polarized intensity detected outside the cores are classified as core-jets with the remaining 10\% being CSO candidates and complex sources.  This implies that relatively speaking, jets are more strongly polarized than cores on average within core-jet systems.  In fact, out of the 92 sources with polarized flux density detected both within and outside the cores, only one source has a value of $f_{pol}$ that is larger for the core than for the regions outside the core.\par
To explore any trend between $f_{pol}$ and source type, we have displayed a so called "box-and-whisker" plot of $f_{pol}$ versus source type in the lower panel of Fig.\ \ref{pfrac}, excluding CPLX sources for which there were only two sources with detected polarized flux density.  These plots indicate that the median, upper and lower quartiles, and extreme values of $f_{pol}$ are all larger for LJET sources than for both point-like and short jet sources, regardless of whether the entire object, the core, or the regions outside the core are used to compute $f_{pol}$.  The median values for CSO sources are similar to those for LJET sources.  One should keep in mind, however, that objects classified as CSO are CSO {\it candidates} and that many (if not all) of these candidates with detected polarized flux density may in fact be core-jet systems.  The results in Table \ref{classes} imply that in fact, detected polarized intensity is relatively rare among the CSO candidate sources with only 15 out of 103 sources having significant polarized flux density.  By comparison, about 40\% of LJET sources have detected polarized flux density, the highest of any source type, which is at least in part due to the relatively large values of $f_{pol}$ found for these sources.
\subsubsection{EVPA and Jet Direction}
\begin{figure}
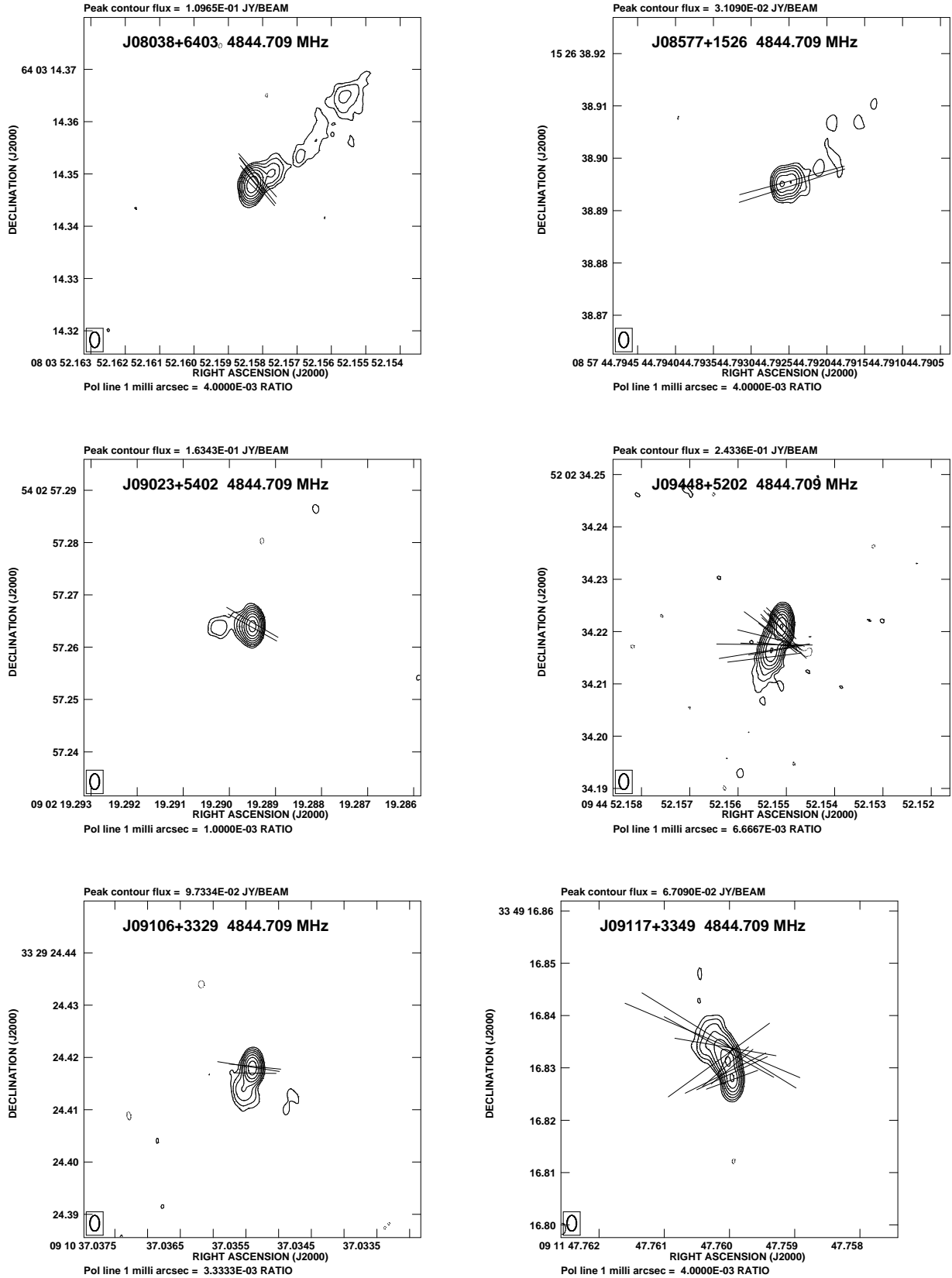

\vspace{-0.35in}
\plottwo{f11a.eps}{f11b.eps}
\plottwo{f11c.eps}{f11d.eps}
\plottwo{f11e.eps}{f11f.eps}
\vspace{-0.35in}
\caption{Examples of six objects identified as core-jets (see \S 2.3) that have detected polarized flux density.  In each plot, the contours correspond to the total intensity and the orientation of the lines is taken from the polarization angle image (see \S 2.4) and represent the EVPA {\it without} any correction for Faraday rotation.  The length of each line represents the ratio of the polarized to total intensity; the scale for these lines is listed below each plot.}
\label{expoljet}
\end{figure}

To explore the possibility of a relationship between the direction of core magnetic fields and the direction of the cores' associated jets, we have computed the absolute difference between the polarization angle and jet position angle of each source with detected polarized flux density classified as a core-jet (see Fig.\ \ref{expoljet} for some examples of polarized sources with jets).  For each of these sources, we take the measurement of the EVPA, $\chi$, outside the core (see \S 2.4) to be the EVPA of the jet component, or $\chi_{jet}$.  For this analysis, we have excluded any source with a single dominant Gaussian component that was classified as a core-jet because it was more elongated than the restoring beam used.  This was done for two reasons.  First, while such objects are too elongated to be unresolved point sources, they have not been resolved into separate components and measuring the polarization properties of their core and jet components separately would be difficult.  Second, since these sources are only marginally resolved, their jet position angles will tend to the position angle of the restoring beam (i.e., PA$_{jet}=0$).  This only effects SJET sources, slightly less than half of which each have a single dominant Gaussian component.\par
\begin{figure}
\plotone{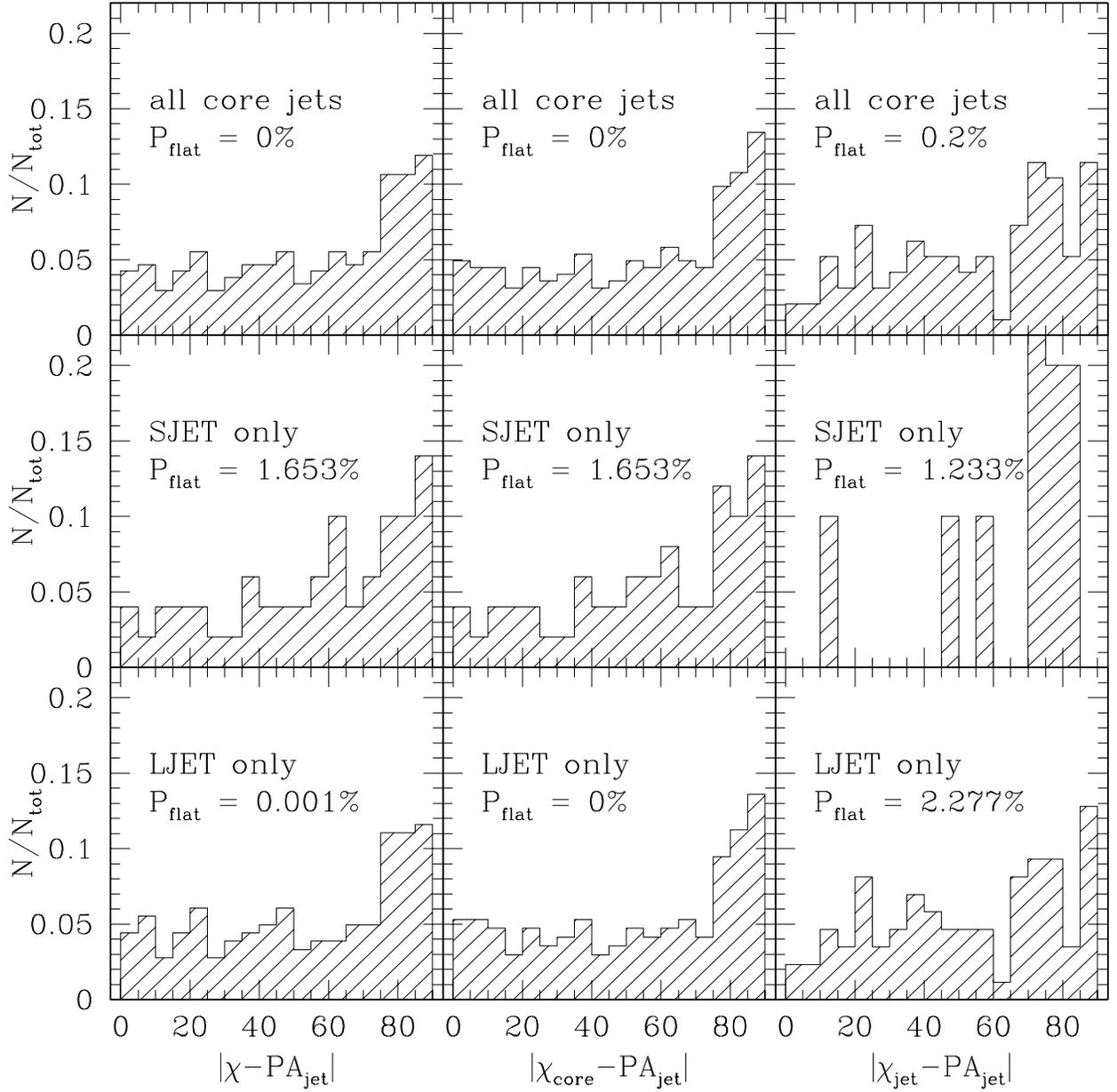}
\caption{The distributions for the absolute difference between the jet position angle and the polarized intensity-weighted mean EVPA, $\chi$, measured for each object (left panels), the core of each object (middle panels), and the jet component of each object which we take to be the regions outside the core (right panels; see \S 2.4).  Distributions are displayed separately for all core-jet systems (upper panels), SJET sources (middle panels), and LJET sources (lower panels).  The probability that each distribution was drawn from a flat distribution computed using a K-S test is displayed in the appropriate panel as $P_{flat}$; values of $P_{flat}<0.001$\% are listed as $P_{flat}=0$.}
\label{pola2}
\end{figure}

 In Fig.\ \ref{pola2}, we have plotted the distributions for $|\chi-\mbox{PA}_{jet}|$, $|\chi_{core}-\mbox{PA}_{jet}|$, and $|\chi_{jet}-\mbox{PA}_{jet}|$ for all core-jet systems and separately for SJET and LJET sources.  For each of these distributions, we have used a K-S test to compute the probability that the distribution was drawn from a flat distribution and have printed the results in the corresponding panels of Fig.\ \ref{pola2} as $P_{flat}$.  For all core-jet and LJET sources, the $|\chi-\mbox{PA}_{jet}|$ and $|\chi_{core}-\mbox{PA}_{jet}|$ distributions are noticeably peaked near 90$^{\circ}$; the probability that each was drawn from a flat distribution is $\leq$0.001\%.  This implies that these peaks are likely the result of a real tendency for the core EVPAs and jet position angles within core-jet systems to be perpendicular to one another which is consistent with what was found for quasars by \citet{pol03}.  The results for the remaining distributions are more marginal but hint that the same is true for SJET sources and that jet position angles may also tend to be anti-aligned with jet EVPAs, but much less frequently than with core EVPAs.\par
\begin{figure}
\plotone{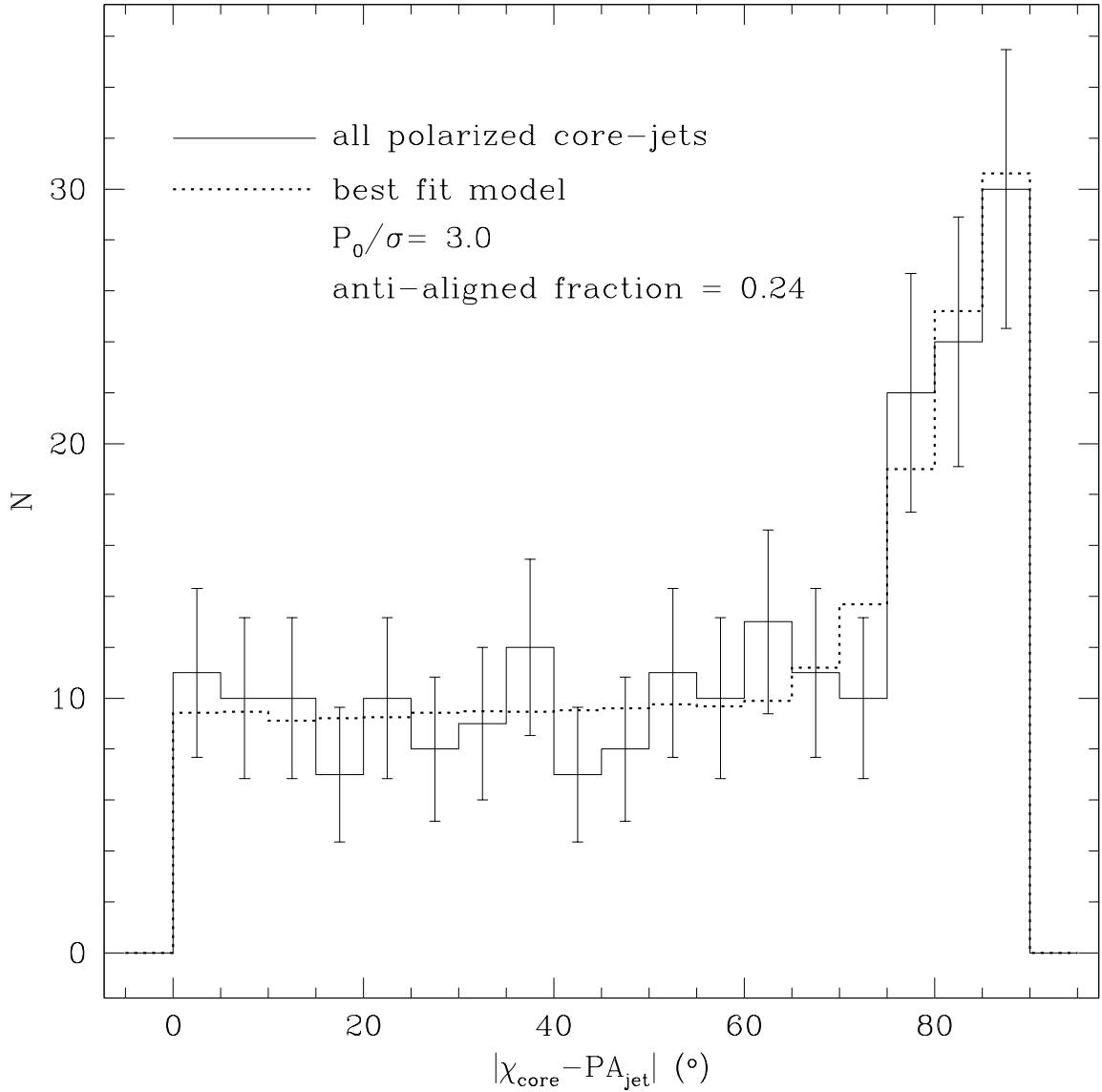}
\caption{The observed (solid line and error bars) and best fitting model (dotted line; see \S 3.2) distributions for the absolute difference between the core EVPA and the jet position angle for core-jet systems with detected polarized flux density.  The best fitting model implies that about 24\% of core-jet systems have EVPAs that are perpendicular to their jet position angles and that the ratio of the uncertainties in the Q and U flux densities for these systems are dominated by Faraday rotation (see \S 3.2).}
\label{pola3}
\end{figure}

To explore the tendency for $\chi_{core}$ and PA$_{jet}$ to be anti-aligned implied by the distribution for all core-jet and LJET sources shown in Fig.\ \ref{pola2}, we have constructed the following simple model.  First, we assume that a significant fraction of polarized core-jet systems have intrinsic polarized flux densities, $P_{0}$, with components $Q_{0}$ and $U_{0}$ which have been altered by a combination of observational errors and Faraday rotation and which have EVPAs that are perpendicular to the jet axes.  We then define a new coordinate system within the (Q,U) plane by rotating the Q and U coordinates such that the new coordinates, $Q^{\prime}$ and $U^{\prime}$, are perpendicular and parallel to the jet axis respectively, i.e., $Q^{\prime}_{0}=0$ and $U^{\prime}_{0}=P_{0}$.  To simulate the influences of observational error and Faraday rotation, we assumed that magnitude of both these effects is the same for the $Q^{\prime}$ and $U^{\prime}$ components of the polarized flux density.  We then assumed that these two effects are additive and that the distribution of the appropriate additive factors can be approximated by a single Gaussian function such that the $Q^{\prime}$ and $U^{\prime}$ components are given by
\begin{eqnarray}
Q^{\prime} = R_{GQ}\sigma \\
U^{\prime} = P_{0} + R_{GU}\sigma
\end{eqnarray}
where $R_{GQ}$ and $R_{GU}$ are two separate random numbers drawn from unit normal distributions and $\sigma$ represents the rms uncertainty in both the $Q^{\prime}$ and $U^{\prime}$ flux densities caused by observational errors and Faraday rotation.  For convenience, we also define a parameter $c=P_{0}/\sigma$ so that we may use two sets of random numbers drawn from unit normal distributions to compute model distributions for $|\chi_{core}-\mbox{PA}_{jet}|_{\perp} = \mbox{tan}^{-1}(U^{\prime}/Q^{\prime})/2$.  We also allow for the possibility that some fraction of core jet systems have core EVPAs and jet position angles that are completely unrelated by computing a second model distribution according to $|\chi_{core}-\mbox{PA}_{jet}|_{random} = \mbox{tan}^{-1}(R_{GU}/R_{GQ})/2$.  Using 10$^{5}$ pairs of random numbers, we used these two model distributions to iteratively solve for the best fitting values of $c$ and the fraction of sources with anti-aligned EVPAs and jet position angles.  The observed and best fitting distributions are plotted in Fig.\ \ref{pola3}; the best fit model is for $c=3.0$ and implies that the fraction of core-jets with anti-aligned EVPAs and jet position angles is 0.24.  An estimate of the covariance matrix for these parameters was computed, and it was found that the two parameters are moderately anti-correlated with a correlation coefficient of $-$0.66.\par
The best fitting value of $c$ implies that on average, $\sigma/P_{0} \simeq$0.33.  For the newly observed VIPS core-jet systems with detected polarized flux density, the median ratio of the rms errors measured from the Q and U images to the peak polarized flux density is $\sim$0.1.  This implies that in order for the model results to be reasonable, Faraday rotation must dominate the uncertainty in the EVPAs, contributing nearly 80\% of the uncertainty in the Q sand U flux densities.  This is not only reasonable, but expected since the RMs of quasars have been observed to be $\sim$500 to a few thousand rad m$^{-2}$ \citep{zav04}, corresponding to rotations of at least $\sim$100$^{\circ}$ at a frequency of 5 GHz.  A proper and more detailed exploration of the relationship between core EVPA and jet position angle will require follow-up observations at multiple frequencies so that RM values may be obtained for our core-jet sources and the appropriate corrections for Faraday rotation can be made.

\subsection{Additional Follow-up}
\begin{figure}
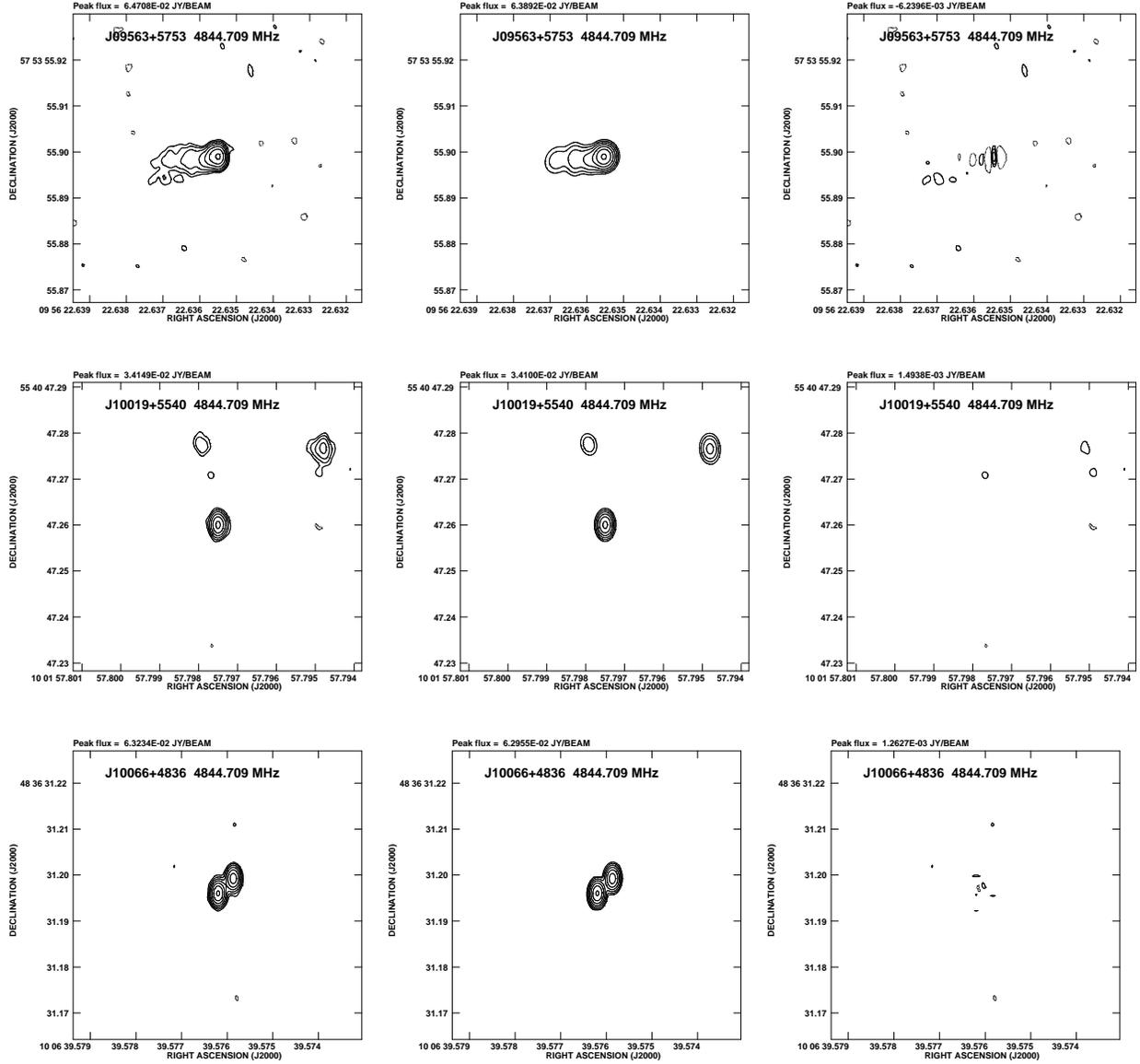

\includegraphics[angle=-90,width=2.1in]{f3a.eps}
\includegraphics[angle=-90,width=2.1in]{f3b.eps}
\includegraphics[angle=-90,width=2.1in]{f3c.eps}
\includegraphics[angle=-90,width=2.1in]{f3d.eps}
\includegraphics[angle=-90,width=2.1in]{f3e.eps}
\includegraphics[angle=-90,width=2.1in]{f3f.eps}
\includegraphics[angle=-90,width=2.1in]{f3g.eps}
\includegraphics[angle=-90,width=2.1in]{f3h.eps}
\includegraphics[angle=-90,width=2.1in]{f3i.eps}
\caption{For three VIPS sources, the 5 GHz I maps (left column), the Gaussian fits to the I maps (middle column), and the residuals for the Gaussian component fits (right column) for a core-jet system (upper panels), a complex system (middle panels), and a CSO candidate (lower panels).}
\label{exgauss}
\end{figure}

With this initial data set, we now have the ability to design follow-up experiments to yield even more information about the nature of AGN.  In particular, with a relatively large sample of good CSO candidates, we are in a good position to use follow-up VLBI observations at other frequencies to confirm whether or not these objects are in fact CSOs and to use multiple epochs to explore how these objects evolve.  Among the CSO candidates and complex sources, we have identified $\sim$20 sources that are good candidates for small separation supermassive binary black hole (SBBH) systems similar to 0402+379 \citep{rod06}.  A good example of one of these sources, J10019+5540, can be seen in the middle panels of Fig.\ \ref{exgauss}.  Ongoing VLBA follow-up observations at 5, 8, and 15 GHz will allow us to confirm whether these systems are indeed compact SBBH systems.  We will also be able to obtain rotation measures for all polarized core-jets ($\sim$300 sources) using follow-up VLBI observations at additional frequencies, allowing for a much more thorough exploration of the relation between magnetic fields and jet activity.  Currently, follow-up observations of all core-jets with detected polarized flux density, all CSO candidates, and all complex sources not included in the ongoing SBBH candidate follow-up program are planned with the VLBA at 5, 8, and 15 GHz.  Finally, future observations of core-jet sources with GLAST will allow us to constrain the physics involved with jets even further.\par

\acknowledgements
The National Radio Astronomy Observatory is a facility of the National Science Foundation operated under cooperative agreement by Associated Universities, Inc.  Information regarding SCHED can be found at \mbox{http://www.aoc.nrao.edu/$\sim$cwalker/sched/sched/sched.html}. The website for the 5 GHz VLA/VLBA polarization calibration database is\\\mbox{http://www.vla.nrao.edu/astro/calib/polar/}.  The authors would like to thank the University of New Mexico and Stanford University for the purchase of Mk 5 disks for the VLBA which was instrumental in achieving a larger data rate, making this survey possible.\\
\indent Funding for the Sloan Digital Sky Survey (SDSS) and SDSS-II has been provided by the Alfred P. Sloan Foundation, the Participating Institutions, the National Science Foundation, the U.S. Department of Energy, the National Aeronautics and Space Administration, the Japanese Monbukagakusho, and the Max Planck Society, and the Higher Education Funding Council for England. The SDSS Web site is http://www.sdss.org/.\\
\indent The SDSS is managed by the Astrophysical Research Consortium (ARC) for the Participating Institutions. The Participating Institutions are the American Museum of Natural History, Astrophysical Institute Potsdam, University of Basel, University of Cambridge, Case Western Reserve University, The University of Chicago, Drexel University, Fermilab, the Institute for Advanced Study, the Japan Participation Group, The Johns Hopkins University, the Joint Institute for Nuclear Astrophysics, the Kavli Institute for Particle Astrophysics and Cosmology, the Korean Scientist Group, the Chinese Academy of Sciences (LAMOST), Los Alamos National Laboratory, the Max-Planck-Institute for Astronomy (MPIA), the Max-Planck-Institute for Astrophysics (MPA), New Mexico State University, Ohio State University, University of Pittsburgh, University of Portsmouth, Princeton University, the United States Naval Observatory, and the University of Washington.

{}
\clearpage
\begin{deluxetable}{r@{$\;\;$}c@{$\;\;$}c@{$\;\;$}c@{$\;\;$}c@{$\;\;$}r@{$\;\;$}r@{$\;\;$}r@{$\;\;$}c@{$\;\;$}c@{$\;\;$}c@{$\;\;$}r@{$\;\;$}r@{$\;\;$}r@{$\;\;$}r@{$\;\;$}r@{$\;\;$}r@{$\;\;$}r@{$\;\;$}r}
\rotate
\tabletypesize{\scriptsize}
\tablecolumns{15}
\tablewidth{0pt}
\tablecaption{Source Properties}
\tablehead{
\colhead{} & \colhead{} & \colhead{$\alpha$ (J2000)} & \colhead{$\delta$ (J2000)} & \colhead{} & \colhead{$F_{8.5}$} & \colhead{$F_{5}$} & \colhead{$F_{5,max}$} & \colhead{rms$_{5}$} & \colhead{} & \colhead{} & \colhead{} & \colhead{$\overline{R}$} & \colhead{$D_{max}$} & \colhead{$PA_{jet}$} \\
\colhead{} & \colhead{Name} & \colhead{($^{h}\:^{m}\:^{s}$)} & \colhead{($^{\circ}\:^{'}\:^{''}$)} & \colhead{UT Date} & \colhead{(mJy)} & \colhead{(mJy)} & \colhead{(mJy/beam)} & \colhead{(mJy/beam)} & \colhead{$N_{GC}$} & \colhead{T$_{\mbox{\scriptsize a}}$} & \colhead{T$_{\mbox{\scriptsize e}}$} & \colhead{(mas)} & \colhead{(mas)} & \colhead{($^{\circ}$)} \\
\colhead{(1)} & \colhead{(2)} & \colhead{(3)} & \colhead{(4)} & \colhead{(5)} & \colhead{(6)} & \colhead{(7)} & \colhead{(8)} & \colhead{(9)} & \colhead{(10)} & \colhead{(11)} & \colhead{(12)} & \colhead{(13)} & \colhead{(14)} & \colhead{(15)} \\}

\startdata
    1 & J07070+6110 & 07:07:00.6167 & +61:10:11.595 & 1998-02-08 &    230.8 &    249.9 &    163.1 &     0.55 &        7 &     LJET &  \nodata &     28.8 &    111.2 &  $-$68.6 \\ 
    2 & J07199+4459 & 07:19:55.5116 & +44:59:06.854 & 2006-04-14 &    180.0 &    156.7 &      8.3 &     0.25 &       10 &     LJET &  \nodata &     20.9 &     68.6 &  $-$21.1 \\ 
    3 & J07260+3912 & 07:26:04.7381 & +39:12:23.335 & 2006-04-14 &    133.0 &     79.1 &     25.1 &     0.22 &        2 &     SJET &  \nodata &      2.9 &      5.8 &  $-$24.0 \\ 
    4 & J07263+4124 & 07:26:22.4226 & +41:24:43.662 & 2006-04-14 &    109.4 &    107.4 &     88.2 &     0.23 &        2 &     LJET &  \nodata &      5.6 &     11.2 &  $-$38.3 \\ 
    5 & J07268+6125 & 07:26:51.6789 & +61:25:13.683 & 2006-05-31 &    110.3 &     99.4 &     86.8 &     0.21 &        1 &       PS &  \nodata &      1.2 &  \nodata &  \nodata \\ 
    6 & J07270+4844 & 07:27:03.1012 & +48:44:10.122 & 2006-04-14 &    263.2 &    263.2 &    224.6 &     0.18 &        2 &     LJET &  \nodata &      3.4 &      6.7 &   $-$8.1 \\ 
    7 & J07288+5701 & 07:28:49.6309 & +57:01:24.375 & 1998-02-08 &    644.3 &    390.5 &    311.7 &     0.96 &        3 &      CSO &  \nodata &      6.7 &     16.4 &  \nodata \\ 
    8 & J07308+4049 & 07:30:51.3491 & +40:49:50.822 & 1998-02-08 &    368.9 &    314.1 &    238.7 &     0.21 &        1 &       PS &  \nodata &      1.2 &  \nodata &  \nodata \\ 
    9 & J07334+5605 & 07:33:28.6148 & +56:05:41.730 & 2006-05-31 &    126.8 &    116.8 &      2.6 &     0.25 &       10 &     LJET &      CSO &     12.5 &     35.9 &  $-$28.3 \\ 
   10 & J07338+5022 & 07:33:52.5222 & +50:22:09.057 & 1996-08-17 &    734.2 &    613.2 &    473.5 &     0.47 &        2 &     SJET &  \nodata &      2.8 &      5.6 &  $-$28.3 \\ 
   11 & J07350+4750 & 07:35:02.3129 & +47:50:08.427 & 1998-02-08 &    460.5 &    438.0 &    302.9 &     0.18 &        2 &     SJET &  \nodata &      2.4 &      4.7 &     88.0 \\ 
   12 & J07359+5925 & 07:35:56.3022 & +59:25:22.128 & 2006-05-31 &     86.5 &     54.6 &     34.5 &     0.24 &        1 &       PS &  \nodata &      1.2 &  \nodata &  \nodata \\ 
   13 & J07362+2954 & 07:36:13.6638 & +29:54:22.198 & 2006-01-27 &    369.9 &    276.7 &    181.8 &     0.24 &        2 &     LJET &  \nodata &      6.9 &     13.8 &     66.7 \\ 
   14 & J07365+2840 & 07:36:31.1975 & +28:40:36.836 & 2006-01-27 &     93.7 &     40.2 &     22.5 &     0.22 &        3 &     LJET &       PS &     31.6 &     71.1 &     21.6 \\ 
   15 & J07369+2604 & 07:36:58.0744 & +26:04:49.888 & 2006-01-27 &    277.2 &    242.3 &    127.8 &     0.29 &        2 &      CSO &  \nodata &      4.2 &      8.5 &  \nodata \\ 
   16 & J07375+5941 & 07:37:30.0858 & +59:41:03.190 & 1998-02-08 &    248.3 &    134.8 &     34.7 &     0.22 &        3 &     LJET &  \nodata &      9.6 &     25.4 &   $-$7.9 \\ 
   17 & J07379+2651 & 07:37:54.9751 & +26:51:47.462 & 2006-05-31 &     87.1 &     72.0 &     53.6 &     0.31 &        2 &     LJET &     SJET &      3.0 &      6.0 &  $-$79.3 \\ 
   18 & J07379+6430 & 07:37:58.9799 & +64:30:43.369 & 2006-05-31 &    239.2 &    356.2 &    147.2 &     0.22 &        2 &     LJET &  \nodata &      3.0 &      6.0 &  $-$50.1 \\ 
   19 & J07395+6306 & 07:39:34.7978 & +63:06:05.570 & 2006-05-31 &     86.5 &     70.7 &     37.9 &     0.21 &        4 &     LJET &  \nodata &      8.5 &     23.6 & $-$166.4 \\ 
   20 & J07398+4423 & 07:39:52.5769 & +44:23:49.692 & 2006-04-14 &    104.1 &     84.5 &     26.2 &     0.25 &        1 &       PS &  \nodata &      1.3 &  \nodata &  \nodata \\ 
\enddata

\tablecomments{Col.\ (1): VIPS source number.  Col.\ (2): VIPS source name.  Col.\ (3): Right ascension (J2000).  Col.\ (4): Declination (J2000).  Col.\ (5): UT date of the observations. Col.\ (6): The flux density at 8.5 GHz from the CLASS survey.  Col.\ (7): The total cleaned flux density from the 5 GHz VLBA map. Col.\ (8): The peak flux density from the 5 GHz VLBA map.  Col.\ (9): The rms noise of the 5 GHz VLBA image.  Col.\ (10): The number of dominant Gaussian components (i.e., that contain more than 95\% of the total flux) fit to the 5 GHz VLBA map (see \S 2.3).  Col.\ (11): The source type derived from the automated Gaussian component classification (see \S 2.3).  Col.\ (12): The source type determined by visual inspection of the I image for sources where the ``by-eye'' and automatic classifications disagree.  Col.\ (13): The mean radius (i.e., mean distances from the mean component position) for the ensamble of dominant Gaussian components.  Col.\ (14): The maximum separation among the dominant Gaussian components.  Col.\ (15): The jet position angle (measured from north through east).} 
\label{props}
\end{deluxetable}

\begin{deluxetable}{lcc}
\tablecolumns{3}
\tablewidth{0pc}
\tablecaption{Summary of Source Types}
\tablehead{
\colhead{Type} & \colhead{N} & \colhead{N$_{pol}$/N} \\
\colhead{(1)} & \colhead{(2)} & \colhead{(3)}}

\startdata
PS & 276 & 31$\pm$4\% \\
SJET & 241 & 36$\pm$5\% \\
LJET & 471 & 41$\pm$4\% \\
CSO & 103 & 15$\pm$4\% \\
CPLX & 17 & 11$\pm$8\% \\
ND & 11 & \nodata \\
\enddata

\tablecomments{Col.\ (1): Source type determined using the ``by-eye'' reclassifications of the automatic classifications where applicable (see \S 2.3).  Col.\ (2): Number of sources within the class listed in Col.\ (1).  Col.\ (3): Percentage of sources with detected ($>5\sigma$) polarized flux within the class listed in Col.\ (1) (not including sources with only CJF or VIPS pilot data).} 
\label{classes}
\end{deluxetable}

\begin{deluxetable}{rcccccccc}
\tablecolumns{9}
\tablewidth{0pt}
\tablecaption{Source Properties}
\tablehead{
\colhead{} & \colhead{} & \multicolumn{3}{c}{$f_{pol}$} & \colhead{} & \multicolumn{3}{c}{$\chi$ ($^{\circ}$)} \\
\cline{3-5}
\cline{7-9}
\colhead{ID} & \colhead{Name} & \colhead{total} & \colhead{core} & \colhead{outside core} & \colhead{} & \colhead{total} & \colhead{core} & \colhead{outside core} \\
\colhead{(1)} & \colhead{(2)} & \colhead{(3)} & \colhead{(4)} & \colhead{(5)} & \colhead{} & \colhead{(6)} & \colhead{(7)} & \colhead{(8)}}

\startdata
    2 & J07199+4459 &    0.296 &    0.263 &    0.797 & &      -53 &      -52 &      -70 \\ 
    4 & J07263+4124 &    0.024 &    0.024 &  \nodata & &       74 &       74 &  \nodata \\ 
    5 & J07268+6125 &    0.027 &    0.027 &  \nodata & &      -13 &      -13 &  \nodata \\ 
    6 & J07270+4844 &    0.120 &    0.085 &    0.370 & &       86 &       85 &      -81 \\ 
   13 & J07362+2954 &    0.019 &    0.019 &  \nodata & &      -59 &      -59 &  \nodata \\ 
   15 & J07369+2604 &    0.019 &    0.019 &  \nodata & &       81 &       81 &  \nodata \\ 
   17 & J07379+2651 &    0.036 &    0.036 &  \nodata & &      -29 &      -29 &  \nodata \\ 
   21 & J07405+2852 &    0.033 &    0.033 &  \nodata & &       89 &       89 &  \nodata \\ 
   25 & J07425+4215 &    0.031 &    0.031 &  \nodata & &       62 &       62 &  \nodata \\ 
   27 & J07426+5444 &    0.023 &    0.023 &  \nodata & &      -33 &      -33 &  \nodata \\ 
   28 & J07431+3941 &    0.022 &    0.022 &  \nodata & &       19 &       19 &  \nodata \\ 
   34 & J07464+2549 &    0.035 &    0.035 &  \nodata & &      -12 &      -12 &  \nodata \\ 
   35 & J07466+2734 &    0.053 &    0.053 &  \nodata & &      -61 &      -61 &  \nodata \\ 
   38 & J07486+2400 &    0.118 &    0.054 &    0.467 & &      -36 &      -39 &        5 \\ 
   41 & J07501+5015 &    0.086 &    0.072 &    0.131 & &      -22 &      -21 &      -24 \\ 
   44 & J07516+2657 &    0.024 &    0.024 &  \nodata & &      -86 &      -86 &  \nodata \\ 
   46 & J07518+3313 &    0.046 &    0.046 &  \nodata & &        4 &        4 &  \nodata \\ 
   54 & J07547+4823 &    0.030 &    0.028 &    0.107 & &      -74 &      -74 &      -65 \\ 
   58 & J07569+5151 &    0.044 &    0.044 &  \nodata & &       15 &       15 &  \nodata \\ 
   64 & J08011+4401 &    0.033 &    0.033 &  \nodata & &       13 &       13 &  \nodata \\ 
\enddata

\tablecomments{Col.\ (1): VIPS source number.  Col.\ (2): VIPS source name.  Col.\ (3): Average 5 GHz fractional polarization (see \S 2.4).  Col.\ (4):  Average fractional polarization within the brightest Gaussian component.  Col.\ (5):  Average fractional polarization outside the brightest Gaussian component.  Col.\ (6):  Polarized intensity-weighted mean electric vector position angle (EVPA) at 5 GHz (see \S 2.4).  Col.\ (7):  Polarized intensity-weighted mean EVPA within the brightest Gaussian component.  Col.\ (8):  Polarized intensity-weighted mean EVPA outside the brightest Gaussian component.}

\label{polprops}
\end{deluxetable}

\clearpage


\clearpage
\thispagestyle{empty}
\setlength{\voffset}{-17mm}
\begin{figure}
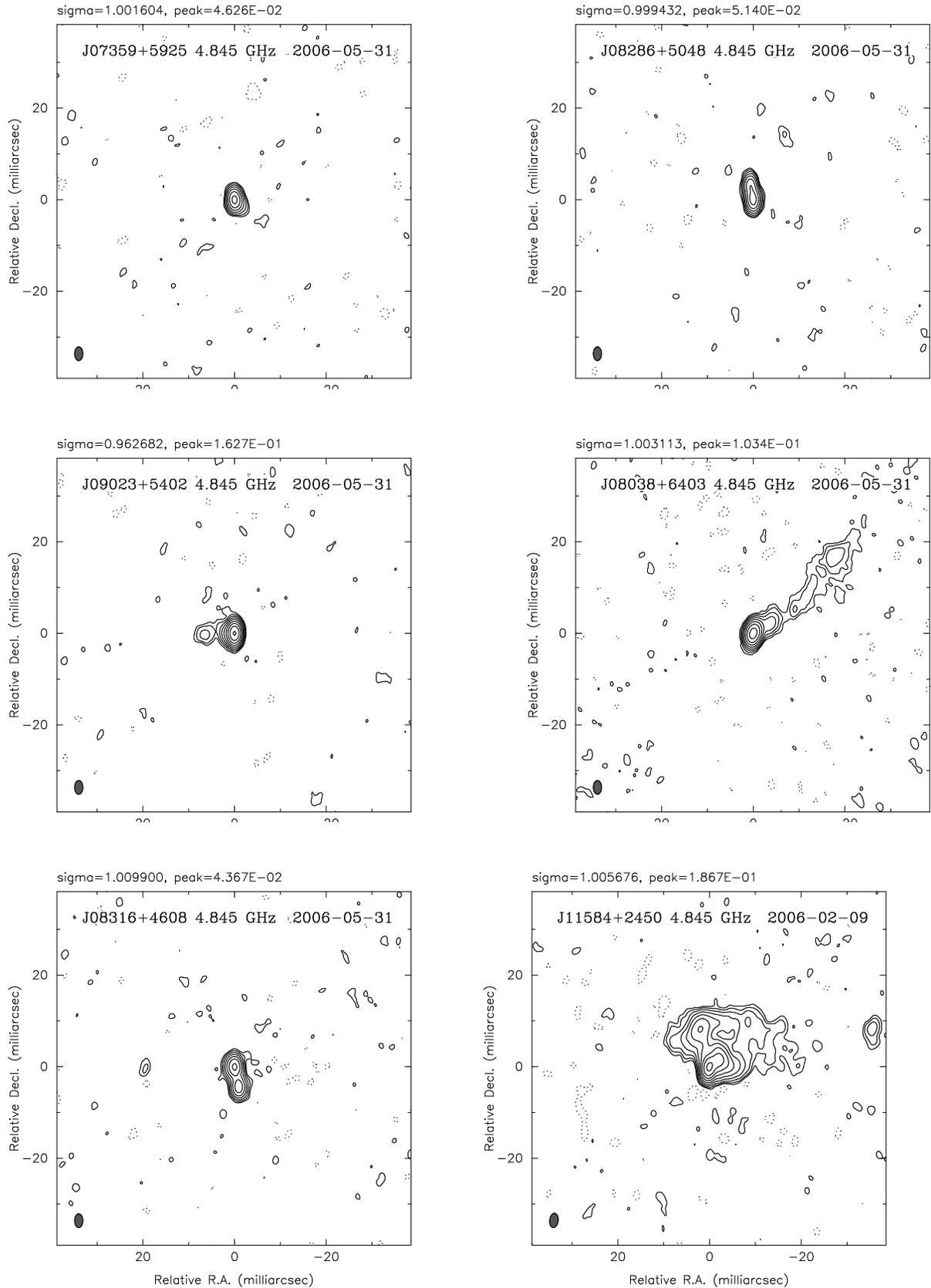

\vspace{-0.5in}
\plottwo{f6a.eps}{f6b.eps}
\vspace{-1in}
\plottwo{f6c.eps}{f6d.eps}
\vspace{-1in}
\plottwo{f6e.eps}{f6f.eps}
\vspace{-1in}
\caption{An example of a point-like object (upper left), a single component core-jet (upper right), a double component core-jet (middle left), a multiple component long jet (middle right), a CSO candidate (bottom left), and a complex source (bottom right) as determined by the automated classification algorithm detailed in \S 2.2.  Here, "sigma" and "peak" refer to the reduced $\chi^{2}$ between the CLEAN components and the data and the peak cleaned flux density in Jy beam$^{-1}$ respectively.}
\label{exclass}
\end{figure}
\clearpage
\setlength{\voffset}{0mm}

\clearpage
\thispagestyle{empty}
\setlength{\voffset}{-17mm}
\clearpage
\setlength{\voffset}{0mm}

\end{document}